\documentclass[aps,pre,twocolumn,showpacs,superscriptaddress,groupedaddress]{revtex4-1}
\usepackage{graphicx}  % needed for figures
\usepackage{dcolumn}   % needed for some tables
\usepackage{bm}        % for math
\usepackage{amssymb} % for math#
\usepackage[utf8]{inputenc}
\usepackage{amsmath,mathtools,amssymb,latexsym}
\usepackage{xspace,colortbl,color}
\usepackage{fullpage,enumerate}
\usepackage{epsfig,amsopn,graphicx}
\def\bea{\begin{eqnarray}}
\def\eea{\end{eqnarray}}

\def\nn{\nonumber}

\def\red#1{\textcolor{red}{#1}}

\def\cH{\mathcal{H}}
\DeclareMathOperator*{\argmax}{\arg\!\max}

\def\texitem#1{\vskip 5pt \par \noindent\hangindent 12pt
  \hbox to 16pt {\hss #1 ~}\ignorespaces}

\newtheorem{lemma}{Lemma}

\begin{document}
%\lipsum
%opening
\title{Cascade of transitions in molecular information theory}
\author{Suman G. Das}
\affiliation{Simons Centre for the Study of Living Machines, National Centre for Biological Sciences (TIFR),
GKVK Campus, Bellary Road, Bangalore 560065, India}
\author{Madan Rao}
\affiliation{Simons Centre for the Study of Living Machines, National Centre for Biological Sciences (TIFR),
GKVK Campus, Bellary Road, Bangalore 560065, India}
\author{Garud Iyengar}
\affiliation{Industrial Engineering and Operations Research, Columbia University, New York, NY 10027}

\begin{abstract}

Biological organisms are open, adaptve systems that can respond to changes in environment in specific ways. 
Adaptation and response can be posed as an optimization problem, with a tradeoff between the benefit 
obtained from a response and the cost of producing environment-specific responses. Using recent results in stochastic thermodynamics, 
we formulate the cost as the mutual information between the environment and the stochastic response. The problem of 
designing an optimally performing network now reduces to a problem in rate distortion theory -- a branch of information theory
that deals with lossy data compression.
We find that as the cost of unit information goes down, the system undergoes a sequence of transitions, corresponding
to the recruitment of an increasing number of responses, thus improving response specificity as well as the net payoff. We derive formal equations for the transition points and exactly solve them for special cases. 
The first transition point, also called the {\it coding transition}, demarcates the boundary between a passive response and 
an active decision-making by the system. We study this transition point in detail, and derive three classes of asymptotic 
behavior, corresponding to the three limiting distributions of the statistics of extreme values. Our work points 
to the necessity of a union between information theory and the theory of adaptive biomolecular networks, in particular 
metabolic networks.

\end{abstract}


\maketitle
\section{Introduction}
Biological systems are distinctive in their ability to adaptively respond
to different environments. This applies not   
only to organisms, but even % single
individual 
cells in multicellular organisms\cite{alberts}. Adaptive behavior contains the notion of 
specificity, i.e. the 
% where a
response % must be
is tailored towards the particular
stimulus~\cite{specificity1,specificity2,bialek}.  % that produces it  
The response results in a benefit
to the cell. For example, % in the context of energy metabolism,
the 
synthesis of enzymes specific to the breakdown of  
a particular nutrient in the environment leads to a free-energy gain for
the cell. The net free-energy payoff should however include, apart from this gain, the free-energy {\it cost} of response
specificity. An optimal adaptive network % that  
% has evolved to perform optimally will
evolves to maximize % the net payoff, which is
the difference between the benefit and  
the specificity cost. 
The degree of specificity in this biological context has been hard to quantify; % part of the
% reason is that
part of the reason is that specificity is not the  
property of  particular pathways, but of the entire repertoire of 
possible cellular responses and their regulatory mechanisms. Recent
developments in information thermodynamics \cite{parrondo} allow  
us to build a coarse-grained model of specificity cost in a system.

\begin{figure}
\includegraphics[height=2.8in,width=3.1in]{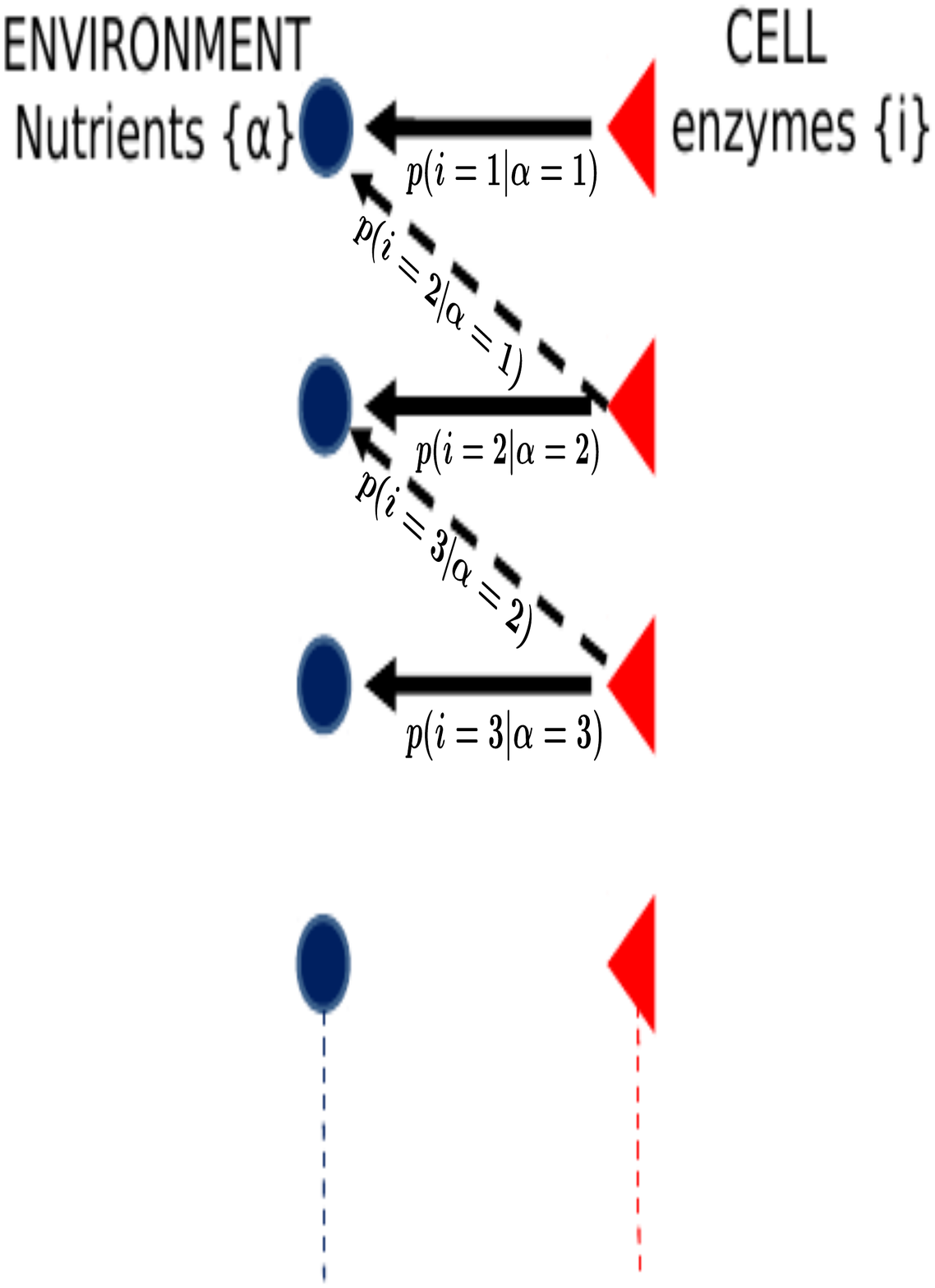}
\caption{A cell (organism) must respond adaptively to different environmental
  states. The environment comprises a diversity of nutrients that 
  occur with frequency $f_\alpha$ ($\alpha=1,\dots,n$). The cell responds to the environment  
  with response probabilities
  $p(i|\alpha)$, where $i$ labels the cellular response. Adaptive behavior is seen as a communication channel  
  between environment and response.}
\label{schematic}
\end{figure}

We model the adaptive response as a problem in
% communication
coding or data compression.  Consider the schematic in \figurename{~\ref{schematic}}. The environment comprises of 
a diversity of states indexed by $\alpha$, and they are presented to the cell with frequencies $f_\alpha$.  % presented with  
% a particular
The cell produces a response $i$ with
probability $p(i\vert \alpha)$. In information theoretic terms, the
environment state $\alpha$ is the signal and the response $i$ is the \emph{code}
or the representation, and the mapping $p(i \vert \alpha)$
is a random code. 
The stochastic response to the environment
models the fact that responses 
at the cellular level are   
inherently noisy. % , and therefore the response must be characterized by a
% response probability distribution $p(i|\alpha)$. 
% In terms of information theory, $p(i|\alpha)$ represents a channel, with
% $\alpha$ as an input and $i$ as the (probabilistic) 
% output \cite{cover}. 
From a bio-chemical perspective, the adaptive response essentially involves a transfer of
information from one set of molecules to another.  
For example, many bacteria such as {\it E. coli} can grow on a wide
variety of sugars \cite{alon}. The concentration of various  
enzymes in a bacterial cell contains information about the nutrients in
its growth medium~\cite{bialek}. 
Since the environmental states occur with frequency $f_\alpha$, the
% nput and output pair satisfy a joint  
% distribution
the joint distribution of the environment state $\alpha$ and the response
$i$ is given by $P(i,\alpha)=p(i|\alpha) f_\alpha$. 
The mutual information between the environment $\alpha$
and reponse $i$ is 
given by  
\begin{eqnarray*}
I
  & = & \sum_\alpha f_\alpha \sum_i p(i|\alpha) \log
        \frac{P(i,\alpha)}{p(i)f(\alpha)},\\
  & = & 
        \sum_\alpha f_\alpha \sum_i p(i|\alpha) \log \frac{p(i|\alpha)}{p(i)},
\end{eqnarray*}
where $p(i)=\sum_\alpha P(i,\alpha)$ is the marginal distribution of the
response. The mutual information $I$ denotes the average number of nats about the
environment state $\alpha$ learned from observing the response
$i$. Thus, the mutual information is high when the residual uncertainty in
the environment state $\alpha$ is low after observing the response $i$,
i.e. the response is very specific to each environment state. 

% This is a measure of the specificity of response, since it is the
% number of bits   
% about the input that can be recovered from the output.  In perfectly
% general terms, the result of a classical measurement of 
% a system with discrete states indexed by $\alpha$ is given by the conditional distribution $p(i|\alpha)$, and $I$ tells us the 
% amount of information gained from the measurement.  
We assume here that when the environment states change to a new randomly
chosen state,  
the adaptive response of the system is instantaneous, and  the memory of
the previous response is erased. This dynamics occurs on a scale much faster than adaptive evolution 
over fitness landscapes.
Recent results in information thermodynamics have established that 
the minimum free energy consumed in such measurement and erasure processes
is proportional  
to the mutual information $I$
\cite{parrondo,tlusty_07,sgm}.
Thus, the cost of specificity % , viewed as a measurement process,
is proportional to $I$.
This cost is of fundamental thermodynamic origin, connected to problems involving the Maxwell's demon \cite{parrondo}, 
and it is also irreducible, so that it cannot be circumvented by better system design.
Here we ignore the cost of producing the response machinery, but focus only on the operational cost of regulating the machinery
to produce the appropriate response.
% It must be stated that
Note that mutual information is 
a lower bound for this cost; the precise value of the free energy cost
depends on the specifics of the system, and may, in general, be
significantly   
greater than this lower bound. Nevertheless, what we lose in sacrificing some of the
biological realism, we gain in the generality and  
tractability of the model. This allows us to identify qualitative regimes of behavior even in the absence of detailed data.
  
Having identified the cost of specificity, we can understand the performance of an adaptive system 
as involving a tradeoff between benefit and cost. We ask what an optimal response distribution, $p(i|\alpha)$, looks like,
given the abundances of various environmental states. We investigate this question by deriving the general features of 
these distributions. Our main focus will be on biological systems that perform energy catabolism, because in these systems  
it is the free energy  cost that is of primary relevance in analyzing system performance. In other systems, such as those 
involving the synthesis of proteins through transcription and translation, it is not obvious how to construct a 
simple qunatitative indicator of performance,
and one should investigate 
whether the free energy cost is indeed an important limiting factor. Nonetheless, the information theoretic 
approach has been suggested as relevant 
for the origin of molecular codes in a variety of contexts \cite{tlusty_07,tlusty_iop}.
In the next section, we state our model for energy catabolism formallly as an optimization problem.
\section{Model}
Consider an organism that is exposed to $n$ different nutrients % free energy
% sources (nutrients)
indexed by $\alpha$, that occur 
in the environment with probability $f_\alpha$. % , and confer a maximum free
% energy benefit of $E_\alpha$.
The organism % , 
% when presented with the source
on sensing environment 
$\alpha$ produces a response (i.e. the synthesis of the appropriate 
transporters, enzymes, etc.) indexed by $i$
with a probability $p(i|\alpha)$. 
Let $E_{i\alpha} \geq 0$ denote the free-energy benefit to the organism from
response $i$ in % the 
% presence of free-energy source 
environment
$\alpha$. 
  %Note that we are \emph{not} assuming
  %the space of responses $i$ is the same as that of the free energy source
%$\alpha$.} % is
% The response needs to be highly
% specific, so that the free energy $E_\alpha$ is 
% extracted only if $i=\alpha$. 
The average free energy benefit is given by  
\begin{equation}
  G=\sum_i \sum_\alpha f_\alpha p(i|\alpha) E_{i\alpha}.
\end{equation}
The free energy cost of sensing the environment and producing a % specific
particular response, i.e. coding for the environment, is
proportional to the mutual information~\cite{parrondo,tlusty_07,sgm,tlusty_iop}.
%\begin{equation}
% I=\sum_\alpha f_\alpha \sum_i p(i|\alpha) \log \big(\frac{p(i|\alpha)}{p(i)}\big),
%\end{equation}
%where $p(i) = \sum_{\alpha} p(i\vert \alpha) f_{\alpha}$. 
The net free energy payoff is therefore 
\begin{equation}
  F=G(p)-\kappa^{-1} I(p),
\end{equation}
where the constant $\kappa$ is determined by general thermodynamic and
chemical parameters that are independent of the channel.  A higher
$\kappa$ is associated with a lower  
cost of specificity.

An optimal metabolic network would tune the channel $p(\cdot|\cdot)$
  to maximize  
the payoff $F$, i.e. solve the optimization problem
\begin{equation}
    \mbox{max}_{p(\cdot,\cdot)}\ G - \kappa^{-1}\ I(p).
    \label{RD}
\end{equation}
Note that this problem is equivalent to the problem
\begin{equation}
  \label{eq:metabolic-2}
    \mbox{min}_{p(\cdot,\cdot)}\ -\kappa \sum_i\sum_{\alpha} f_{\alpha}
    p(i\vert \alpha) E_{i\alpha}  + I(p),
\end{equation}
i.e. a rate distortion problem with the distortion function $d(i,\alpha) =
- E_{i\alpha}$ \cite{cover}. The distortion function is a loss term
that measures the performance of the code, wheareas $I$ is the quantity of information 
with which to minimize the loss.
%information about the source contained 
%in the code. 
%Rate distortion theory seeks to minimize the loss given a constraint on $I$.
In our interpretation, the first term in \eqref{RD} is a benefit term that depends on the specificity of 
the molecular code, wheareas the fidelity incurs a cost given by the second term. 
% The optimal channel is a function of the parameter set
%$\{\{f_\alpha,E_\alpha\},\kappa\}$. 

This formulation is not limited to metabolic responses. It should be
applicable wherever the free energy cost is relevant,  
and the benefit is linear in the conditional probabilities. For example,
rate distortion theory has been used to investigate  
the origins of the genetic code in \cite{tlusty_07}. 

\begin{figure}
\hspace{.3cm} \includegraphics[height=1.8in,width=2.75in]{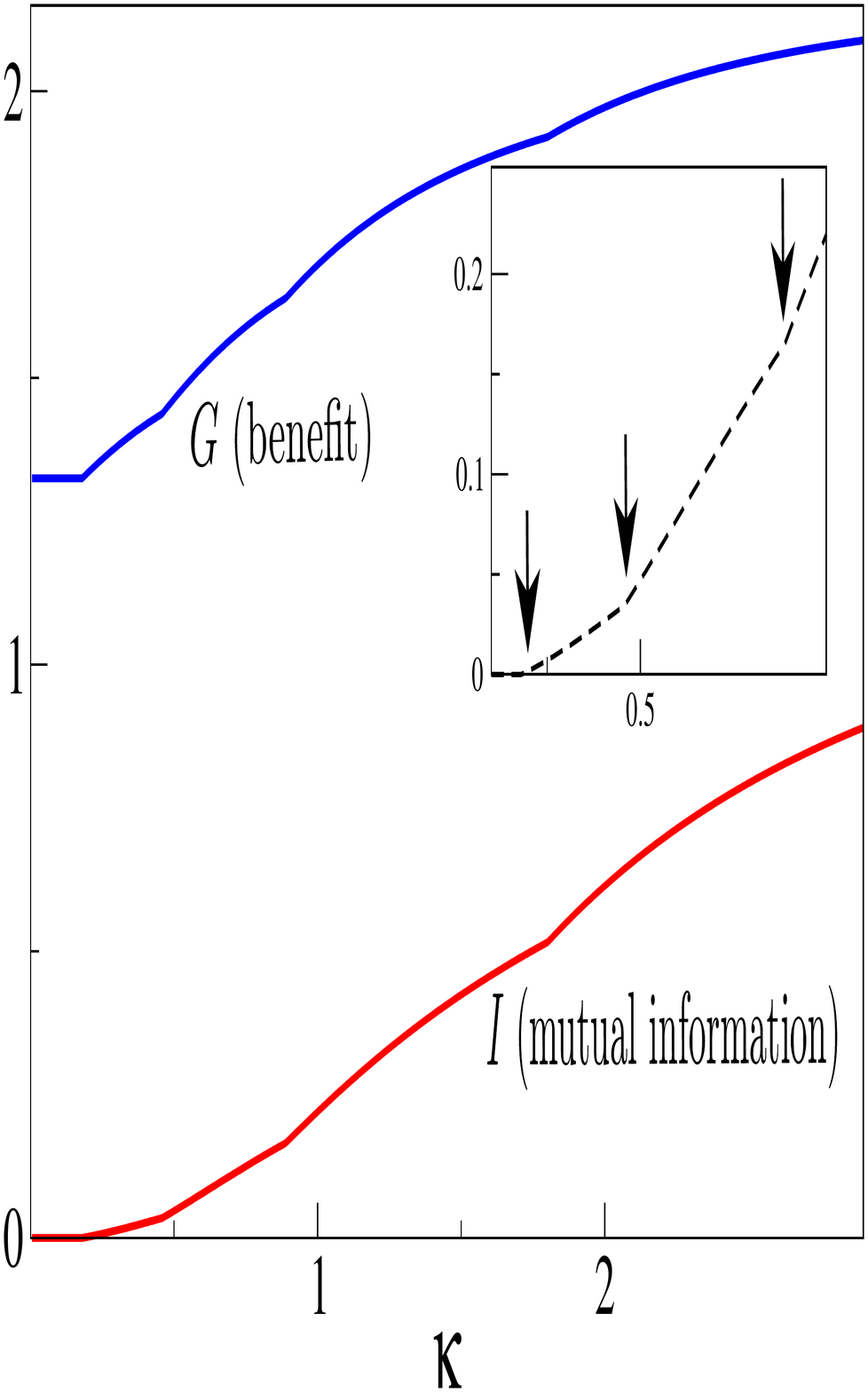}
 \includegraphics[height=1.7in,width=2.4in]{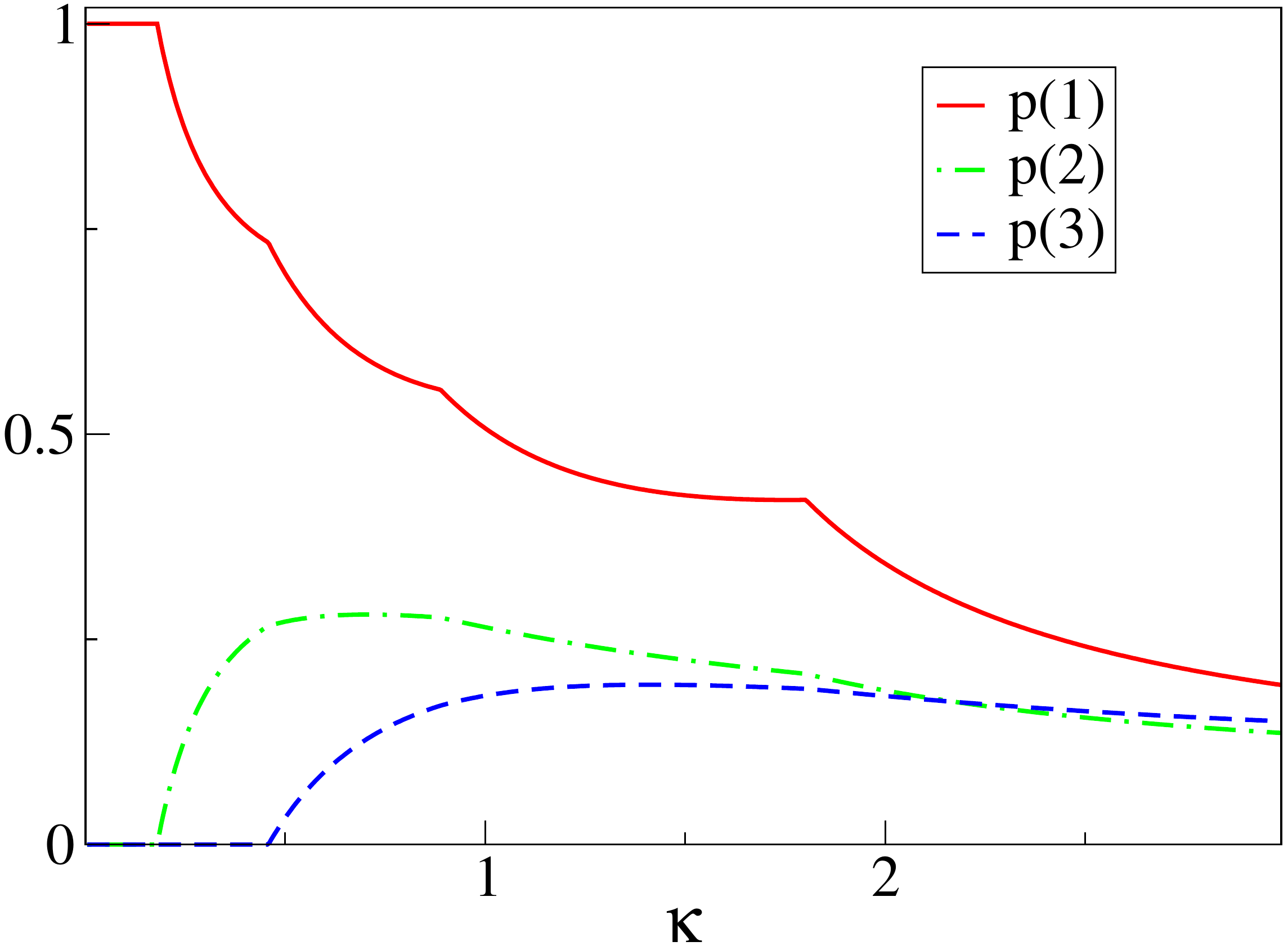}
\caption{Cascade of transitions with $n=6$. Parameters: $E_{\alpha,\alpha}=\alpha+0.2$,
  $f_\alpha \sim 6/(\alpha-0.4)$,  
 $E(i,j)=0.2 E(i,i)$, for $j \neq i$. (Upper panel) The benefit and mutual
 information as a function of $\kappa$ are plotted. 
 The inset 2
 zooms in on the first three transition points, marked by arrows. (Lower panel) The first three responses to be turned
 on. The response $p(1)$ is the optimal non-coding response; $p(2)$ is turned on at the first transition point, also called 
 the coding transition.
}
 \label{bif}
\end{figure}

%Our goal in this paper is to ivestigate the behavior 
%of the optimal channel and the mutual information as a function of $\kappa$ and the nutrient distribution.
As a generic example of how the optimal solution behaves, we numerically compute in \figurename{~\ref{bif}} the
mutual information and conditional probabilities  
for a particular parameter set when $n=6$, as a function of $\kappa$.
 The data are generated by calculating the optimal channel by an alternating 
minimization procedure called the Blahut-Arimoto algorithm \cite{cover}. The procedure 
is described in Appendix A.
% The alogrithm used
% to generate the solution is described in  
% Appendix A. 
The notable feature of the plot is that the mutual information $I$ is $0$ at
$\kappa=0$, and $I$ undergoes  
a sequence of bifurcations as $\kappa$
increases, analogous to continuous transitions in thermodynamic
systems. The first bifurcation point  
has been called the coding transition in a related context
\cite{tlusty_iop} -- this is the value of $\kappa$ subsequent to which the mutual
information becomes non-zero. 
Each transition point is characterized by the 
fact that a new response is ``switched on''.  For example, in \figurename{~\ref{bif}}, 
the response probability $p(2)$ becomes non-zero at the coding
transition point. As $\kappa$ 
increases, we see a sequence of such transitions, where typically only one
response is activated at a time.   
We call this sequence of transitions a cascade. Similar cascades have been
observed previously at the interface of statistical mechanics  
and computer science, such as in the application of deterministic
annealing to clustering, classification and other  
computational problems~\cite{rose}.  
The following section finds analytical solutions for the cascades. 

\section{Analysis of cascades}
When $\kappa=0$, the cost of unit information is infinite, therefore the
mutual information $I=0$. Thus, the response $p(i\vert \alpha)$ is, in
fact, independent of the
environment, i.e. $p(i\vert \alpha) = p(i)$. % i.e the response is
% independent of the 
% environment state.
Since $I(p) = 0$, the optimization problem reduces to  
\[
\max_{p}\ \sum_ip(i) \Big(\sum_{\alpha} f_{\alpha}  E_{i\alpha}\Big).
\]
It is clear that the optimal $p^\ast(i) = \delta_{ir}$, where $r =
\text{argmax} \{\sum_{\alpha} f_{\alpha}
E_{i\alpha}\}$. Thus, the organism uses the response $r$ 
irrespective of the environment state $\alpha$.  
We will refer to $r$ as the optimal non-coding response.

As $\kappa$ increases, at some point the mutual information becomes
non-zero, i.e. the response $p(i\vert \alpha)$ \emph{does} depend on
$\alpha$. {This is called a 
coding transition in a related context~\cite{tlusty_iop} because the
organism senses and codes the environment in terms of the response.}
As $\kappa$ continues to increase, we see more and more responses being
activated, each accompanied by a kink  
in the mutual information and the benefit $G$ (see \figurename{~\ref{bif}}). In the simplest interpretation, 
each response represents the expression of a particular gene (or operon). 
The task of regulating the expression is carried out by transcription
factors in the cells, and the free energy cost  
associated with the specificity of the responses arises out of the
activity of the regulatory machinery. The existence  
of cascades tells us that in an optimal metabolic network, a particular
metabolic response may not be expressed  
even when the appropriate enzyme-coding genes 
and the corresponding nutrients are present in the system. 
%When the distribution of nutrients in the environment changes, 
%the same system may produce a response that it previously did not. 
The number of responses available from an 
optimal metabolic network is limited by the unit cost of information.

The distinctive feature of the cascades is that the responses are
activated one at a time. It is not clear {\it a priori} 
that it should be so. One may think naively that for any $\kappa>0$, all responses would be active;
or that, in general, multiple responses should be switched on at a transition point. 
We demonstrate in the following subsection that cascades are indeed generic.  % is what one should generically expect.

\subsection{Cascades are generic}
% A closed expression for the solution of the general rate distortion problem is not
% known. But the criteria for  
% a channel to be optimal are known.
We show below that under generic choice of parameters, two 
different responses cannot transition at the same $\kappa$.
We start by assuming that the vector of free-energy $E_i =
\{E_{i\alpha}\}_{\alpha=1}^n$ corresponding to response $i$ are
distributed according to a distribution that has an $n$-dimensional
density. Thus, $\mathbb{P}(E_i \in A) = 0$ for any set $A$ that has
dimension at most $(n-1)$.

Let $p(i) = \sum_{\alpha} p(i\vert \alpha) f_{\alpha}$ denote the marginal
probability for response $i$. For $\kappa > 0$, results in Chapter 10 of~\cite{cover}
establish that $p$ is optimal, if and only if,
\begin{equation}
  \label{eq:metabolic-3}
  \sum_{\alpha} \frac{f_{\alpha} e^{\kappa E_{i\alpha}}}{\sum_j p(j)
    e^{\kappa E_{j\alpha}}} \left\{
  \begin{aligned}
    \leq 1 && p(i) = 0,\\
    = 1 && p(i) > 0.
  \end{aligned}
  \right.
\end{equation}
Strong convexity of mutual information implies that the optimal
solution $p_{\kappa}$ is unique and a continuous function of $\kappa$. 
The optimal solution for $\kappa=0$ can be obtained by taking the limit of
$\kappa \searrow 0$. %  , the above condition has to be evalued in the
% limit of small $\kappa$.  

Now, fix a response index $s$. We call $\kappa_s$ a transition point
for the response $s$ if $p_{\kappa_s}(s) = 0$ and there exists
$\epsilon_s>0$ such that $p_{\kappa}(s) > 0$ for all $\kappa \in (\kappa_s,
\kappa_s + \epsilon_s]$. By the continuity property it follows that 
\[
\sum_{\alpha} \frac{f_{\alpha} e^{\kappa_s E_{s\alpha}}}{J_{\kappa_s}(\alpha)}=1
\]
where $J_{\kappa}(\alpha) = \sum_j p_{\kappa}(j)
e^{\kappa E_{j\alpha}}$. 

Suppose $\kappa_s$ is a transition point for another index $t \neq
s$, i.e. $p_{\kappa_s}(t) = 0$ and there exists an $\epsilon_t > 0$ such
that $p_{\kappa}(t) > 0$ for all $\kappa \in (\kappa_s, \kappa_s +
\epsilon_t]$. Thus, a necessary condition is that 
\[
\sum_{\alpha} \frac{f_{\alpha} e^{\kappa_s E_{t\alpha}}}{J_{\kappa_s}(\alpha)}=1.
\]
Note that $p_{\kappa_s}(t) = 0$ implies that $E_t= \{E_{t\alpha}\}_{\alpha= 1}^n$ is
not included in the sum defining $\{J_{\kappa_s}(\alpha)\}_{\alpha = 1}^n$. 
%Thus, the vector ${E_t}$ lies in an $(n-1)$ dimensional hyperplane. 
Therefore,
\begin{eqnarray*}
  \lefteqn{\mathbb{P}\big(\text{$\kappa_s$ is a transition point for
  $t$}\big)}\\
  & \leq & \mathbb{P}\left( \sum_{\alpha} \frac{f_{\alpha} e^{\kappa_s
           E_{t\alpha}}}{J_{\kappa_s}(\alpha)}= 1 \right) = 0,
\end{eqnarray*}
where the last equality follows from the fact that the probability that
$E_t$ lies in an $(n-1)$ dimensional hyperplane is zero, which follows from our starting assumption.
(We reiterate that the case of $\kappa=0$ has to be analysed as a limit, in which case the same conclusions will follow.)

\subsection{Structure of bifurcation points for diagonal response}
Starting in this section we will focus on the special case where the
benefits are diagonal, i.e $E_{i \alpha}= E_\alpha \delta_{i \alpha}$. 
This is a reasonable approximation in the context of metabolic responses,
since enzymes are generally highly specific to  
substrates.  In this section, we focus on computing the  bifurcation
points~$\kappa$. % in the system.  
% We introduce the simplification that

Define $g_\alpha=f_\alpha E_\alpha$.
%Assume that a bifurcation happens at some $\kappa^\ast$, with two
%alternative scenarios: \\ 
Then, the object to be maximized is 
\[
 \mathcal{L} = G- \kappa^{-1} I +\sum_\alpha \lambda_\alpha \sum_i p(i|\alpha),
\]
where the last term, involving the set of Lagrange multipliers $\lambda_\alpha$, ensure the normalization 
 of the response probabilities.
Equating the derivative of this to zero, we have 
\begin{equation}
 p(i|\alpha) = p(i) \exp[\kappa (\frac{\lambda_\alpha}{f_\alpha}+\frac{g_i}{f_i} \delta_{i,\alpha})].
 \label{casc0}
\end{equation}
Solving for $\lambda_\alpha$ and substituting the result, we obtain the relations: 
\begin{eqnarray}
  p(i|\alpha) &=& \frac{p(i) e^{\kappa E_i \delta_{i \alpha}}}{1+p(\alpha) (e^{\kappa E_\alpha}-1)}
 % p(i|i) &=& \frac{p(i) e^{k E_i}}{1+p(i) (e^{\kappa E_i}-1)}.
\end{eqnarray}
We use the relation $p(i)= \sum_\alpha p(i|\alpha) f_\alpha$ in
conjunction with the above equations to obtain  
\begin{eqnarray}
  \sum\limits_{\alpha \neq i} \frac{f_\alpha}{1+p(\alpha) (e^{\kappa E_\alpha}-1)}
  +\frac{f_i e^{\kappa E_i}}{1+p(i)(e^{\kappa E_i}-1)} = 1 \nn, 
\end{eqnarray}
for $p(i)>0$.

\begin{figure}
 \includegraphics[height=2in,width=2.5in]{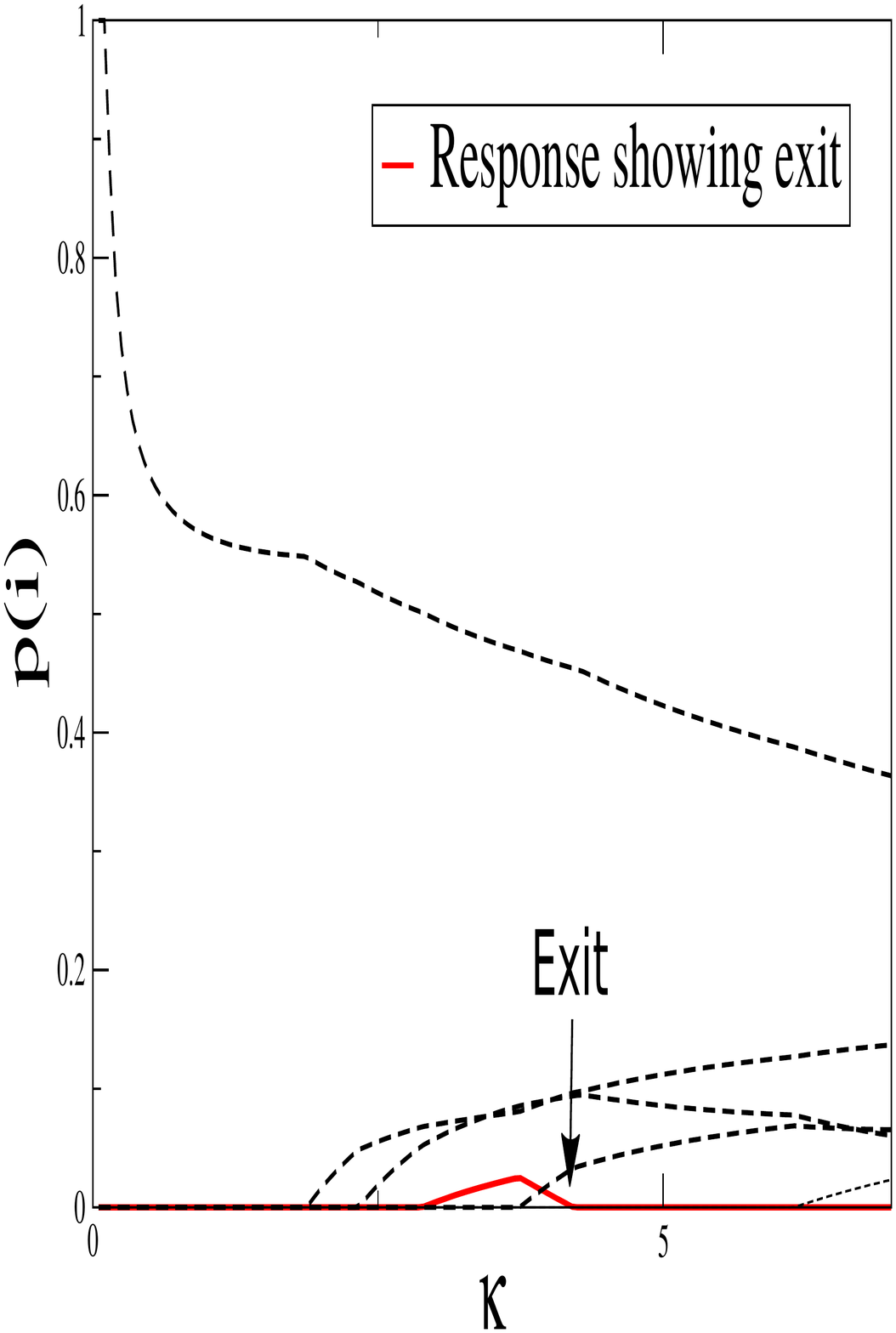}
 \caption{Here we show a simulation of the optimal marginal probabilities $p(i)$, for $n=40$; the diagonal benefit values 
 $E_{ii}$ were 
 chosen randomly from a uniform distribution, whereas the off-diagonal $E_{i\alpha}$ were chosen uniformly from 
 $[0,\sigma_{od} E_{i,i}]$, with $\sigma_{od}=0.9$. The red curve shows a response that is switched on at a particular 
 value of $\kappa$ 
 and exits the active set at a higher value of $\kappa$. 
 Some of the other responses that do not exit the active set are shown 
 in dashed lines for comparison. As $\sigma_{od}$ decreases, instances of exits from the active set become rare.
 }
 \label{exitgraph}
\end{figure}

The sum can be rewritten as 
\begin{eqnarray}
  \sum\limits_{\alpha} \frac{f_\alpha}{1+p(\alpha) (e^{\kappa E_\alpha}-1)}
  +\frac{f_i (e^{\kappa E_i}-1)}{1+p(i)(e^{\kappa E_i}-1)} = 1 \nn
\end{eqnarray}
Define $C=\sum\limits_{\alpha} \frac{f_\alpha}{1+p(\alpha) (e^{\kappa
    E_\alpha}-1)}$. Then, from the above  
equation, 
\begin{equation}
  p(i) = \frac{f_i}{1-C} - \frac{1}{e^{\kappa E_i} -1}.
  \label{ex1}
\end{equation}
Thus, it follows that 
\begin{equation}
  \label{eq:ex-2}
  p(i) > 0 \quad \Leftrightarrow \quad h_i(\kappa) := f_i(e^{\kappa E_i} - 1) > 1-C.
\end{equation}
Let us define the active set as $A = \{i: p(i) > 0\}$. Then substituting \eqref{ex1} in the definition
of $C$, we get 
that 
\[
  C=(1-C) \sum\limits_{i \in A} \frac{1}{e^{\kappa E_i}-1} + \sum\limits_{j \notin A} f_j,
\]
Substituting the solution for $C$ in \eqref{ex1}
we obtain that for all $i \in A$,   
\begin{equation}
  p(i) = \frac{f_i}{\sum\limits_{m \in A} f_m}\big(1+ \sum\limits_{j
    \in A} \frac{1}{e^{\kappa E_j}-1}\big) 
  -\frac{1}{e^{\kappa {E_i}} - 1}.
  \label{ex2}
\end{equation}
This does not specify the solution for the marginals of the response
probability completely,  
since we still need to know the active set $A$. We need to understand how the active set changes as $\kappa$ increases. 
At $\kappa=0$, the only member of the active set is the optimal non-coding response. As $\kappa$ increases, we expect 
the size of the active set to increase. However, it is also possible for responses to exit the active set; 
see \figurename{~\ref{exitgraph}} for such an example. Such cases make the composition of the active set difficult to track and 
the the bifurcation points harder to determine. Below we find conditions under which the behavior of the active set is more regular,
and use it to find the bifurcation points analytically.

First, we fix $\kappa$ and  reorder the indices such that $h_i(\kappa) \geq h_j(\kappa)$
for $i \leq j$. Then \eqref{eq:ex-2} implies that the active set $A = \{i: i
\leq \ell\}$, i.e. we only need to search over $\ell$. The index $\ell$ is
optimal for $\kappa$ if $p_i > 0$ for all $i \leq \ell$, and $p_i=0$ for $i>l$.

Next, we characterize special cases where there exist thresholds $\{\kappa_i: 1
\leq i \leq n\}$ such that $p(i) > 0$ for all $\kappa > \kappa_i$. In other words, no response exits 
the active set as $\kappa$ increases, and the cascade consists entirely of transitions into the active set.

% The transition point $\kappa$
% for the response $i$ is given 
% by 
% \begin{equation}
%  \frac{f_i}{\sum\limits_{l \in A_{\ell}} f_l} \big(1+ \sum\limits_{j \in
%    A_{\ell}} \frac{1}{e^{\kappa E_j}-1}\big) 
%   -\frac{1}{e^{\kappa {E_i}} - 1}=0,
%   \label{eq:trans}
% \end{equation}
% There are instances where we can find 
% the active set independently, and solve Eq.(\ref{ex2}) for the bifurcation points.

% From the approach given here, we can prove some further properties of
% active sets. For example, once a response  enters the active set, can it
% exit the active set as $\kappa$ increases?  We can prove the following
% proposition.
\begin{lemma}
  Suppose either (a) $E_i \equiv E$ or (b) $f_i \equiv f = \frac{1}{n}$. Then
  there exists thresholds $\kappa_m$ such that $p(m) > 0$ for all $\kappa >
  \kappa_m$. 
\end{lemma}
From \eqref{eq:ex-2} it follows that in both the special cases (a) and
(b), the ordering $\{h_i(\kappa): 1\leq i \leq n\}$ is independent of
$\kappa$. Order the indices $i$ such that $h_i(\kappa) \geq h_j(\kappa)$
for all $i < j$.  From \eqref{ex2}, it follows that $p(m) > 0$ if, and only if, 
\[
  \sum_{j \leq m} f_j\left( \frac{1}{h_i(\kappa)} -
    \frac{1}{h_j(\kappa)} \right) < 1
\]
Define $\Delta_i(\kappa) = \frac{1}{h_{i+1}(\kappa)} -
\frac{1}{h_i(\kappa)}$ for $i < n$. We show that in both cases (a) and
(b), $\Delta_i(\kappa) = \frac{1}{h_{i+1}(\kappa)} -
\frac{1}{h_i(\kappa)}$ which is clearly monotonically decreases with $\kappa$. Define
$\kappa_m$ to be the solution of the equation 
\[
  \sum_{j \leq m} f_j\left( \frac{1}{h_i(\kappa)} -
    \frac{1}{h_j(\kappa)} \right)\\
=   \sum_{j < m} \big(\sum_{k \leq j} f_k\big) \Delta_j(\kappa) = 1. 
\]
Then the monotonicity of $\Delta_j(\kappa)$ implies that $p(m)~=~0$ for
all $\kappa \leq \kappa_m$ and $p(m) > 0$ for all $\kappa > \kappa_m$.

For case (a) 
\[
  \Delta_i(\kappa) = \left(\frac{1}{f_{i+1}} - \frac{1}{f_i} \right)
  (e^{\kappa E}-1)^{-1},
\]
and clearly monotonically decreasing in $\kappa$.   
Define $x={(e^{\kappa_m E}-1)}^{-1}$, and $F_m=\sum_{j\leq m} f_j$.
From (\ref{eq:ex-2}) we have
\[
 \frac{f_m}{F_m} (1+ mx)-x=0.
\]
Thus,
\[
 \kappa_m  = \frac{1}{E} \ln \left(\frac{F_m}{f_{m}}-(m-1)\right).
\]
The organism transitions to a coding mode when at least two responses have
a positive probability of occurring. This transition occurs at 
\[
 \kappa_2 =\ln(\frac{f_1}{f_2}).
\]
Thus, the coding transition point depends rather weakly on the ratio
of highest to second highest frequencies. 
% The mutual information $I_m$ at $\kappa_m$ is given by 
% \begin{eqnarray*}
%   \lefteqn{I_m = } \\
% && H(f)- \Big[ \sum_{\beta >m } f_\beta \ln \frac{1}{f_\beta}+(m-1)f_m \ln \frac{1}{f_m}\nn \\
% &&  \mbox{} + (F_m-(m-1)f_m) 
%    \ln \big(\frac{1}{F_m -(m-1)f_m}\big) \Big], \nn
% \end{eqnarray*}
% where $H(f)= - \sum f_\alpha \ln f_\alpha$ is the entropy of the
% distribution of nutrient frequencies. 
% The second term is positive and gives the conditional entropy of the nutrients. 
% To solve a special case of this, let us consider exponentially decaying
% frequencies,
Let us treat some particular instances of this case.
Consider exponentially decreasing frequencies of the form $f_m \sim 2^{-c_n m}$, where $c_n$
is some constant that, in general, depends on the total number of nutrients $n$. 
%First, consider the case where $c_n$
%is independent of $n$; we see that the ratio of any two frequencies is now independent of $n$, and therefore 
%$F_m/f_m$ is independent of $n$. In that case, $\kappa_m$ is independent of $m$. 
The solution for the transition points is: 
\[
 \kappa_m=\ln \big[\frac{2^{c_n}(2^{c_n m}-1)}{2^{c_n}-1}-(m-1)\big]
\]

\begin{figure}
 \includegraphics[height=2in,width=3in]{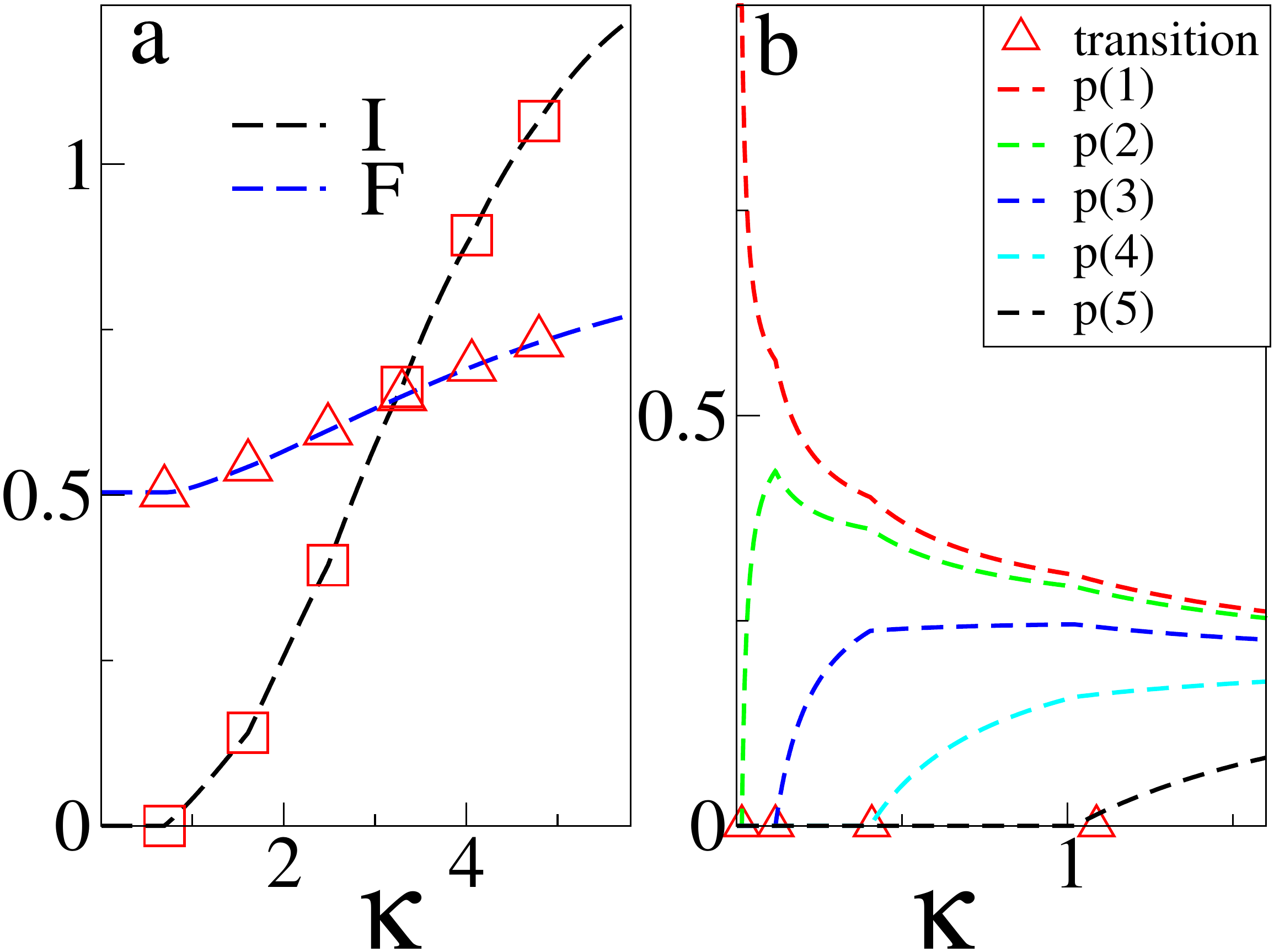}
\caption{ {\bf a.} Mutual information and net payoff for an exactly solvable case, with $E_m=1$, and 
$f_m= \frac{2^{-m}}{1-2^{-n}}$, 
for $n=7$. The dashed 
 lines are from the numerically obatined solutions, whereas the symbols are the exact solutions at the transition point.
{\bf b.} The approximate transition points for the case $f_i=1/n$, $E_\alpha=1-(\alpha-1)^2/n^2$, for $n=8$, are marked as 
triangles. They are compared with the numerically obtained transition points for the second to fifth responses. The approximation 
fits the data well up to the third transition point.}
 \label{analytical}
 \end{figure}

Consider the case where $c_n=1$. 
Then,
% and $m=1,..,n$. Then 
\[
 \kappa_m=\ln (2^m-m).
\]
At large $m$, the transition point $k_m$ is linear in $m$, although the frequencies
decay exponentially in $m$. Thus, % it does not require a very
the information cost need not be very low for the responses to rare
nutrients to be active. This feature can be generically expected because
of the logarithmic dependence of the transition point on
$\frac{F_m}{f_m}$. A numerical verification of the result above is given in \figurename{\ref{analytical}a}.% As an illustration of this point,  
The fact that $\kappa_m$ is independent of $n$ is somewhat counterintuitive, 
since we would expect the transitions to occur earlier when the diversity is larger. 
This would occur if we choose a $c_n$ that goes to zero asymptotically, in which case, for large $n$,
\[
 \kappa_m \approx \frac{c_n}{2}(m-1).
\]
In particular, if $c_n=\frac{1}{n}$, then $\kappa_m \approx \frac{m-1}{2n}$, so the transition points are inversely 
proportional to the diversity $n$. The crucial diffetrence between the two cases is that in the former, the ratio of the 
frequencies are independent of $n$, whereas in the latter case, the different frequencies start to 
crowd together, i.e the ratio of any two frequencies $f_l$ and $f_m$ 
approach $1$ in the limit of large $n$.
%\[
% \kappa_m=\ln \big[\frac{2^{\frac{1}{n}}}{2^{\frac{1}{n}}}(2^{\frac{m}{n}}-1)-(m-1) \big]
%\]

\iffalse
Now consider 
a different case, where $l_n=2^\frac{1}{n}+1$;
In this case, 
\[
 \kappa_m = 
\]
\fi
For case~(b), the derivative
\[
\Delta_i'(\kappa) = \frac{E_i e^{\kappa E_i}}{f(e^{\kappa E_i}-1)^2} -
\frac{E_{i+1} e^{\kappa E_{i+1}}}{f(e^{\kappa E_{i+1}}-1)^2} 
\]
Define $\alpha = \frac{E_{i}}{E_{i+1}} > 1$ and $x = e^{\kappa
  E_{i+1}}>1$. Then 
\[
\Delta_i'(\kappa) = \frac{\alpha x^{\alpha}}{f(x^{\alpha}-1)^2} -
\frac{x}{f(x-1)^2}.
\]
Fix $x > 1$, and define $g(\alpha) = \frac{\alpha
  x^{\alpha}}{(x^{\alpha}-1)^2}$.  
  Then 
  \[
   g^\prime(\alpha)= \frac{x^\alpha W(\alpha \ln(x))}{(x^\alpha-1)^3}
  \]
  where $W(y)=e^y(1-y)-(1+y)$ for $y \geq 0$. Since $x>1$  and $\alpha >
  1$, the sign of
  $g^\prime(\alpha)$ is the same as that of  
  $W$. We see that $W(0)=0$, and 
  $\frac{dW}{dy}=-(y e^y+1) < 0$ for all $y$. Therefore $W(y) < 0$ for
  $y>0$; consequently, $g^\prime(\alpha) < 0$ and  $\Delta_i(\kappa)' =
  \frac{1}{f}\big(g(\alpha) - g(1)\big) < 0$, 
i.e. $\Delta_i(\kappa)$ is monotonically decreasing. 
  % \iffalse
%   \begin{eqnarray*}
%   \ln(g(\alpha))' 
%   & = & \frac{1}{\alpha} + \ln(x) -
%         \frac{2x^{\alpha}\ln(x)}{x^{\alpha}-1}\\
%   & = & \frac{1}{\alpha}\Big( 1 -
%         \big(\frac{x^{\alpha}-1}{x^{\alpha}-1}\big)\ln(x^{\alpha})\Big) 
% \end{eqnarray*}
% Let $w(y) = \big(\frac{1+y}{y-1}\big)\ln(y)$ for $y > 1$. Then $\lim_{y
%   \downarrow 1} w(y) = 1$. We will next establish that $w'(y) \geq 0$. 
% The derivative
% \[
% w'(y) = \frac{1}{y(y-1)^1}\big( y^2  - 1 - 2y\ln(y) \big).
% \]
% Let $u(y) = 2y^2 - 1 - 2y\ln(y)$. Then $u(1) = 0$, and $u'(1) = 0$, and
% $u''(y) = 2 - 2/y \geq 0$ for all $y \geq 1$. Thus, it follows that $u(y)
% \geq 0$ for all $y \geq 0$. Thus, it follows that $w(y) \geq 2$ for all $y
% \geq 1$. Consequently,
% \[
% \ln g(\alpha)' < -\frac{1}{\alpha} \leq 0.
% \]
% \fi

\begin{figure}
 \includegraphics[width=2.8in,height=2.6in]{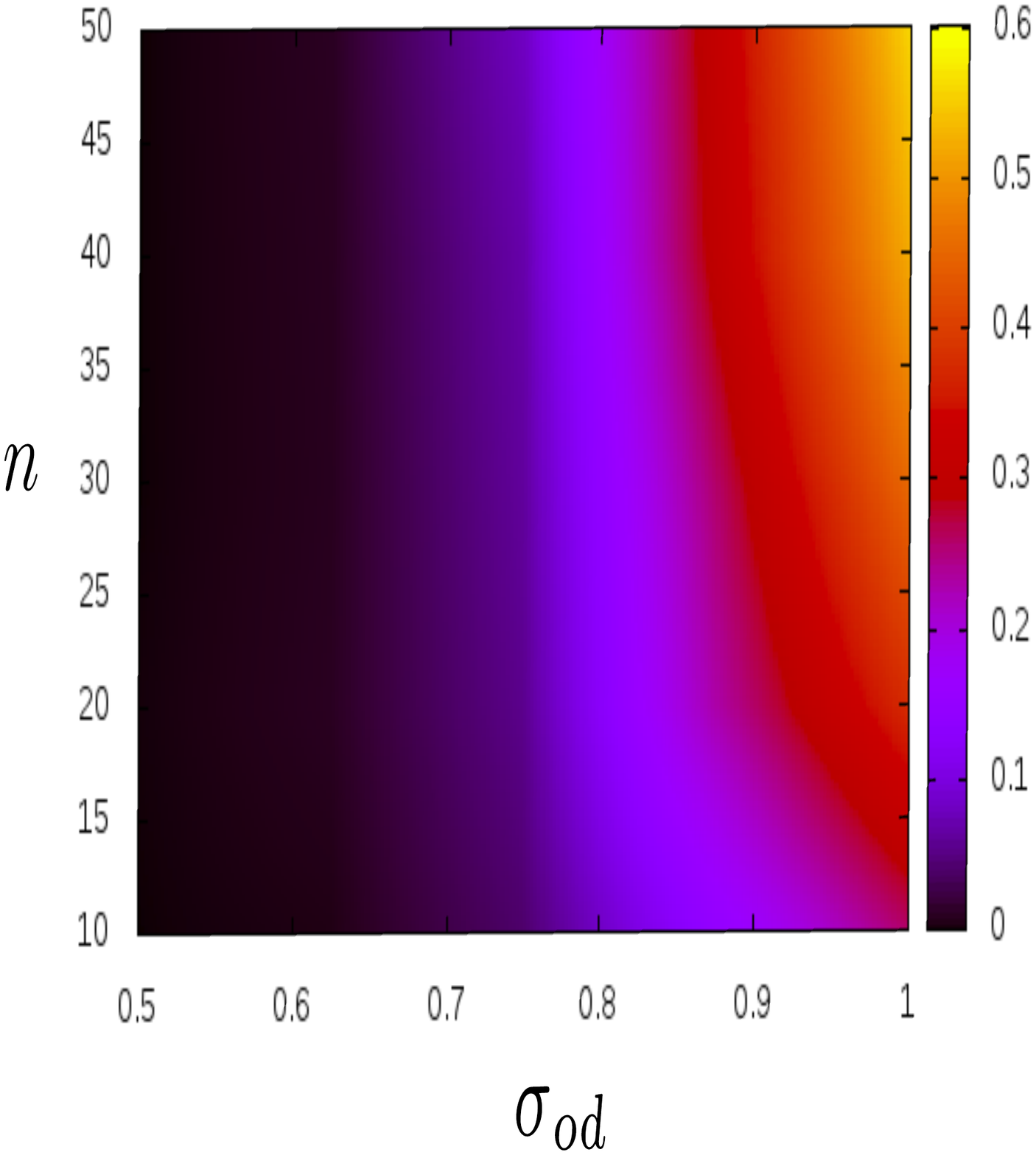}
 \caption{The fraction of cases where at least one exit from the active set happens as $\kappa$ increases from $0$. 
 The diagonal benefit values $E_{ii}$ and the unnormalized frequencies are chosen uniformly from $[0.5,1.5]$. The 
 off-diagonal benefit values $E_{ij}$ are chosen uniformly from $[0,\sigma_{od} E_{ii}]$. When $\sigma_{od}$ is less than 
 $0.5$, exits become extremely rare, and typically no exits are found in a set of $10^3$ different realizations.
 }
 \label{exit}
\end{figure}

In this case, one cannot solve for the bifurcation points $\kappa_m$
exactly. However, one can approximate $\kappa_m$ when 
$\kappa_m E_i \ll 1$ for all $i \leq m$.   
%The last condition would hold, for example, when the benefit values are chosen independently from a distribution that has 
%a sufficiently fast tail, such as a distribution with a finite support (as will be clearer in the next section). 
We can then 
expand \eqref{ex2} up to linear order in $\kappa E_i$ and obtain the solution
\[
 \kappa_m \approx \frac{m}{E_{m}} - \sum_{j=1}^m \frac{1}{E_j}.
\]
See  \figurename{\ref{analytical}b} for comparison of the prediction with data.  
Unlike in Case~(a), the bifurcation points depend more sensitively on the
benefit values $E_i$. Once again, consider  
the coding transition point, $\kappa_2$. If $E_2 \ll E_1$, it can be shown that $\kappa_2 \approx \ln(2)(\frac{E_1}{E_2})$, 
i.e the coding transition depends linearly rather than logarithmically on
the ratio of highest to second-highest benefits.  
In an approximate sense, a low-frequency nutrient is more likely to be
coded for than a low-benefit nutrient. 

In general, the bifurcation points can be solved for when the ordering of the $h_i$ are independent of $\kappa$. This 
happens when $f_i \le f_j$ if and only if $E_i \le E_j$ for all $i,j$. When this condition is violated, the ordering depends on 
$\kappa$, and the 
active set becomes harder to identify. When one also allows for the off-diagonal elememts $E_{i \alpha}$ to be non-zero, 
we find numerically that exits from the active set do happen. It is shown in \figurename{~\ref{exit}} that 
exits are common when the off-diagonal terms are large, but as they become smaller, exits become very rare and 
eventually cannot be numerically detected at all. 

The optimal $p(i\vert \alpha)$ for $\alpha \ne i$ satisfies
\begin{equation}
 \frac{p(i \vert \alpha)}{p(i)} = \frac{1}{1+p(\alpha)(e^{\kappa E_\alpha}-1)} <
 p(\alpha \vert \alpha)
 \label{casc0}
\end{equation}
% A number of properties are worth noting. First, every off-digaonal
% response probablity is less than or equal to the 
% corresponding marginal
% probability, and therefore less than or equal to the diagonal response probability.
This makes intuitive sense, since the 
off-diagonal responses do not contribute to the benefit term. Note that
the off-diagonal $p(i \vert \alpha)$ is positive 
whenever a response is active, i.e. $p(i)>0$. In an experimental context, such
non-cognate responses may show up merely as   
noise, but our analysis here indicates that it may be the result of
an optimization process. Although the off-diagonal  
terms do not increase the benefit, they % can reduce the
contribute to decreasing the 
mutual information,
and therefore increase the net payoff. 
Third, when the response $\alpha$ is not active, we have $p(i|\alpha)=p(i)$, i.e the metabolic response contains no 
information about the nutrients in the non-active set.

Among
all the transition points, the coding transition  
is special because it separates the regime of constant response from a
regime where the organism must sense the environment.
We devote the next section to the analysis of the coding transition,
i.e. the value $\kappa^\ast$ when the conditional distribution $p(i\vert
\alpha)$ is no longer uniform for all environment states $\alpha$. 

\section{Asymptotics of the Coding Transition and the theory of extreme  
  values} 
% As we have seen before, when
% dependent decisions are taken  
% by a cell. 
The coding transition has been suggested as the origin of
molecular codes in biology \cite{tlusty_iop}.  
Here we ask the question: how generic is the phenomenon of metabolic
coding?  When the number of nutrients is  
small, $\kappa^\ast$ has a finite value, but as the number of
nutrients increases, there is an increasing  
benefit to be had from distinguishing the nutrients, and we could expect
$\kappa^\ast$ to be small. But does it go to  
$0$ asymptotically, and if so, at what rate?
We will see in the following that the answer depends on 
the distribution of frequencies and benefits of the nutrients, and certain
universality classes of the coding behavior  
can be identified from the theory of extreme value statistics.

Recall that when 
$\kappa=0$, the optimal response is non-coding, and the 
 optimization problem becomes: 
\[
\max_{p} \sum_i p(i) g_i,
\]
where $g_i =f_i E_i$.
The optimal solution is $p_0(i) = \delta_{ir}$, where $r =
\text{argmax} \{g_i\}$. This solution remains optimal provided
\[
\sum_{\alpha} \frac{f_{\alpha}e^{\kappa E_{i\alpha}}}{\sum_j p_0(j)
  e^{\kappa E_{j\alpha}}} \leq 1
\]
for all $i \neq r$. For the special case where $E_{i\alpha} = E_{\alpha}
\delta_{i\alpha}$, this condition reduces to 
% The coding transition occurs at $\kappa = \kappa^\ast$
\[
 f_i e^{\kappa E_i} + f_r e^{-\kappa E_r} \leq f_i + f_r,
\]
for all $i \neq r$. Thus, the coding transition point $\kappa^\ast$ is
smallest positive value of $\kappa$ that solves one of the  
$n-1$ equations
\[
 \frac{f_i}{f_r} (e^{\kappa E_i}-1)= 1- e^{-\kappa E_r}
\]
for $i \neq r$. Let $x = e^{\kappa E_r}$, $s_i = \frac{E_i}{E_r}$, and
$w_i = \frac{f_i}{f_r}$.  % Then
% without loss of generality, we can put $E_r$ =1, and $\kappa$ is taken
% to be expressed in the units of ${E_r}^{-1}$,   
% and $E_i$ is expressed in units of $E_r$. 
Then the above equation reduces to 
\begin{equation}
 w_i x^{1+s_i}=(1+w_i)x-1.
 \label{eq:wi}
\end{equation}
% where $w_i=\frac{f_i}{f_r}$, and $x=e^\kappa$ and $\rho_i=E_i$. Note that $\kappa$ and $E_i$ are now dimensionless.
One solution is $x^\ast=1$ corresponding to $\kappa=0$. Our goal is to compute
the non-trivial solution $x^\ast > 1$.  
% In real scenarios, the 
In typical scenarios, the precise distribution of nutrient benefits 
$E_{\alpha}$ are unlikely to be known, and hence the precise value of
$x^\ast$ cannot be computed. However, one may be able to identify certain
universal features of the coding transition $x^\ast$ as a function of the
distribution of benefits $\{E_{\alpha}\}_{\alpha = 1}^n$ and the
frequencies $\{f_{\alpha}\}_{\alpha = 1}^n$. 
We assume that $E_\alpha$ are independent identically distributed samples
from a fixed distribution $\pi_E$, and 
$f_{\alpha} = \frac{h_{\alpha}}{\sum_{\alpha^\prime} h(\alpha')}$ where
each term $h_{\alpha}$ is an independent sample from the distribution
$\pi_h$. 

% But under some reasonable 
% assumptions, we can extract a few 
% universal features of the transition point. In fact, the asymptotoic
% scaling of $\rho$ with $n$ 
% can be showed to belong to some universality classes according to 
% the theory of extreme value statistics.
% In general, let the $n$ benefit values be drawn independently from the distribution $Q(x)$.
% For distributions with faster than power law tails, $\kappa \to 0$ as $n \to \infty$, and we can make a useful 
% approximation as follows. 
We will focus on cases where the coding transition occurs at
$\kappa^\ast \approx 0$ as $n \rightarrow \infty$, equivalently, at
$x^\ast = 1 + y$, for $y \ll 
1$. Substituting $y_i=x_i -1$ in  
\eqref{eq:wi}, and expanding up to second order in   
$y_i$, we get 
\begin{equation}
 y_i= \frac{2(1-s_i w_i)}{1-\frac{1}{2} s_i w_i (1-s_i)},
\label{eq:smally}
 \end{equation}
which holds for all $y_i \ll 1$. Since $s_i w_i = \frac{f_i E_i}{f_rE_r} = \frac{g_i}{g_r}<1$, we have that
$y_i>0$, as required. 
% The solution $y$ can be small in the following two different scenarios:
% \begin{eqnarray}
%   s w & \lesssim & 1 \nn\\
%   s & \gg & 1/s w \label{eq:s_gg_1}
% \end{eqnarray}
% %The second option implies that $E_r \gg E_i$.  
We show in the Appendix
that % under fairly general 
% sufficient conditions
when the distribution of $E$ is bounded, the above equation further reduces to 
% , the second condition does not hold; and further, 
% when the first condition holds, we have $s_i \to 1$. 
% in the Appendix 
% However, under fairly general conditions, 
%the  second condition cannot
%hold. % gives sufficient conditions that rule  
% this out, in which case we end up with
% We end up with 
\begin{equation}
  \label{eq:s-cond}
 y_i \approx {2(1-s_i w_i)},
\end{equation}
and therefore the transition corresponds to the largest value of $s_i w_i$.
Under this approximation, the coding transition is determined by the
distribution of the product $g_i = f_i E_i$, % i.e the 
% product of $f_i$ and $E_i$ that matters,  
and not the distributions of the individual terms $f_i$ and $E_i$. 
% and the individual distributions are of no consequence. So let us define a
% new variable 
Define $\rho_i=g_i/g_r$, where $g_r$  
is the benefit associated with the non-coding response. Clearly, $\rho_i < 1$,  and
\begin{equation}
 \kappa^\ast=2(1-\rho),
\end{equation}
where $\rho = \max_{i \neq r} \{\rho_i\}$. %  is the value of $\rho_i$ maximized
% over all $i \neq r$
Thus, the distribution of $\kappa^\ast$ is determined by the distribution
of $\rho$, i.e. 
% Further, 
% we want to find the mean value of $\kappa^\ast$, where the average is
% taken over the distribution from which the  
% $g_i$ are chosen. Thus we need to find
% the mean of the quantity $\rho$. The problem now reduces to one in the
% statistics of extreme values, where we need to  
% find the mean of 
the ratio of the second highest to highest value out of
$n$ values chosen independently from an identical  
distribution $Q(x)$. 
%and for the rest of this section we will denote the mean itself 
%simply as $\rho$, without using any additional notation to indicate averages.
%We are interested in the distribution
%\[
%P_r (\rho)=\int P(x_{max}=x,x_{2max}=\rho x) dx 
%\]
%To evaluate this, we first ask for the joint ditribution $P_2(x_1,x)$ of
%the two highest values,  
%$x$ and $x_1$, $(x \ge x_1)$.
Along the lines of Eq 13 in \cite{satya}, we can write 
\[
 P_2(x,x_1) = n Q(x) P_{max} (x_1;n-1) \Theta(x-x_1)
\]
where the term $n Q(x_1)$ is the probability that {\it any} one of the $n$
chosen numbers has the value $x_1$, and  
$P_{max} ( x;n-1)$ is the probability that among the remaining $n-1$
numbers (which were   
chosen independently from $Q(x)$), $x$ is the highest value; the Heaviside
step function $\Theta$ ensures that  
the probability is zero when $x>x_1$. We need to characterize the ratio
$\rho=x/x_1$ 
using this distribution. It is natural to split 
the problem into three different classes of distribution $Q$, which are
known to exhibit different  
extreme value behavior \cite{book}. 

\subsection{Asymptotic classes}
Suppose $n$ values $X_i$ are sampled independently from a distribution
$Q$. Let $x_n = \max_{1\leq i \leq n} \{X_i\}$ denote the maximum of these
$n$ samples. Then 
% and the
% largest value is $x$,   
there exist % scaling
coefficients $a_n$, and $b_n$ such that the distribution of 
$x_n$ for large $n$ is given by $b_n^{-1} f(\frac{x-a_n}{b_n})$, where the
function $f$ is one of three universal   
limiting forms of extreme value statistics. The precise form the limiting
distribution $f$ is determined by the tail of the
parent distribution $Q$. In this section, we compute the asymptotic behavior
of the value of the coding transition $\kappa^\ast$ for particular distrubutions belonging to each of the three
canonical classes \cite{satya}. We consider the classes in decreasing order of how fast the tail 
of the parent distribution decays.

% Below we consider standard examples from the three classes, and find the
% asymptotic behavior of $\kappa^\ast$. 
\vskip 5 pt
\noindent{\bf (i) Bounded distributions}\
Consider a bounded distribution of the form $Q(x) \sim (a-x)^{\beta-1}$,
$\beta>0$ with the support $0 \le x\le a$. This covers a broad range of
asymptotic behaviors that commonly occur as $x \to a$  
\cite{satya}. In this case, the asymptotic distribution $f(z)$ of the
scaled maximum $z = b_n^{-1}(x_n - 
a_n)$ is of the Weibull form 
\[
  f(z)\sim{(-z)}^{\beta-1} \exp\big(-(-z)^\beta\big)\ 
\] 
for $a_n = a$ and $b_n = n^{-1/\beta}$.
%  and $Q(x)=0$  
% for $x > a$. Also, $\beta>0$. 

\begin{figure}
 \includegraphics[width=2.5in,height=1.8in]{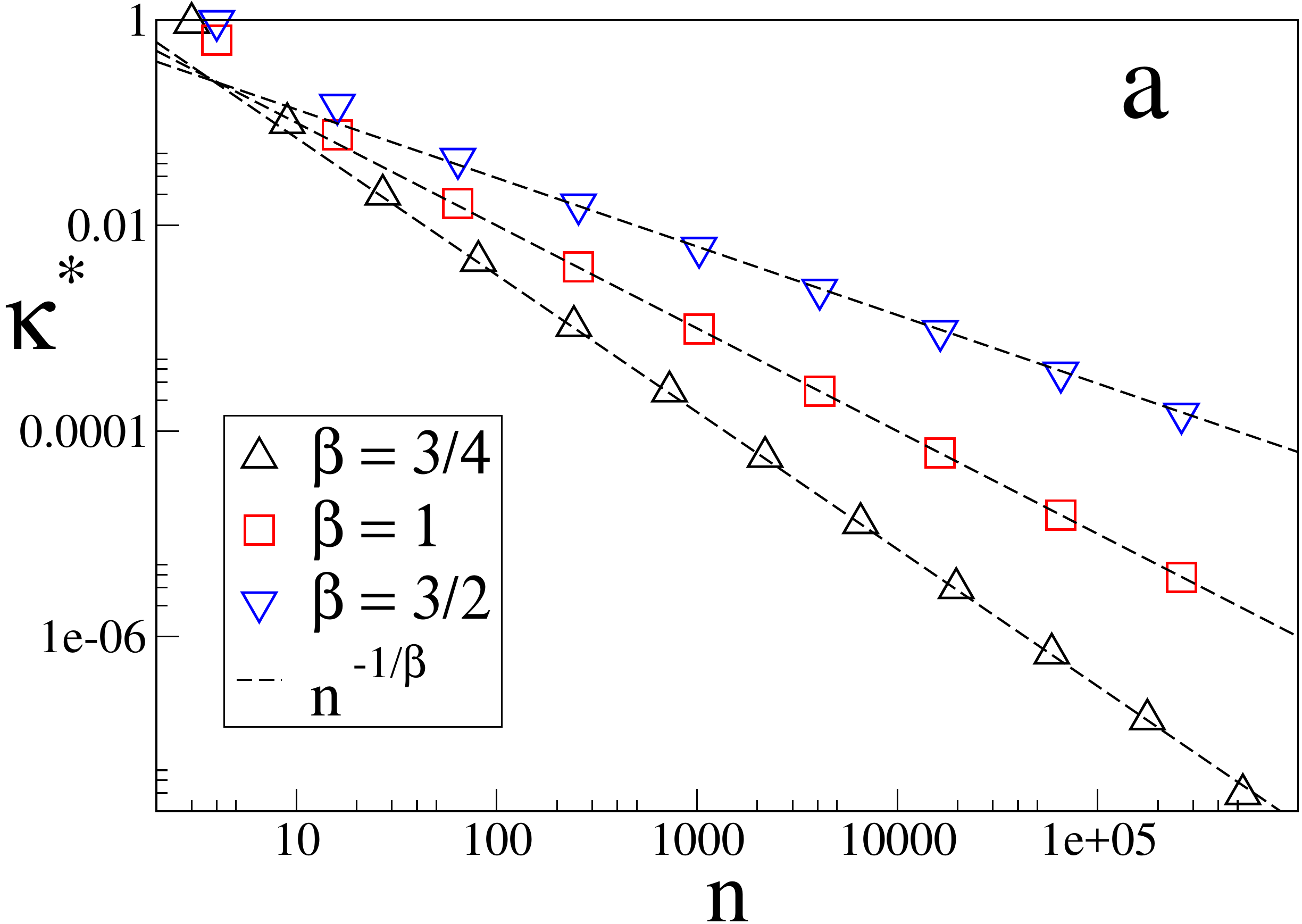}
 
  \hskip 10 pt \includegraphics[width=2.1in,height=2.7in,angle=-90]{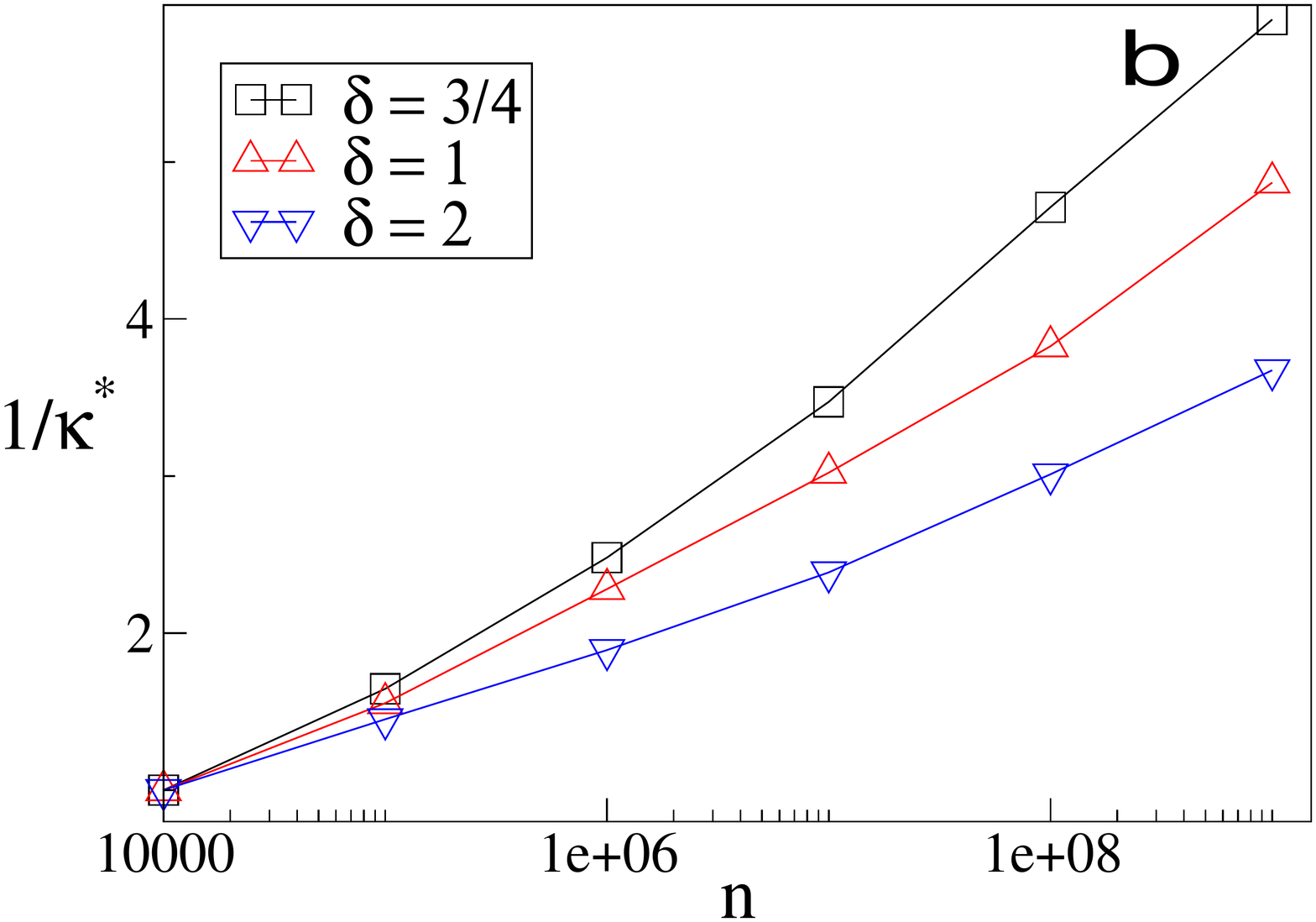}
 \caption{The scaling of the coding transition point as a function of total number of nutrients. 
 (a) The result for bounded distribution, with the predicted scaling shown in dashed lines. The symbols represent data points 
 obtained from numerical simulation. (b) The inverse of $\kappa^\ast$ plotted on log-linear scale. The curves have been 
 normalized by the value at $n=10^4$ for clear visibility in a single frame. 
 The theory predicts 
 a linear curve; but the convergence to the asymptotic logarithmic regime is extremely slow, and only the approach to the 
 regime can be detected in the plots.
 }
 \label{scaling}
\end{figure}

Let $x$ denote second highest value among the $n$ terms, and let $x +
\epsilon$, $\epsilon > 0$, denote the highest value among the $n$
terms. Then the joint distribution of $(x,\epsilon)$ is given by 
% The joint distribution of the second highest value $x$, and the difference
% between the highest and second highest values  
% $\epsilon$ is given by 
$n Q(x+\epsilon) P_{max}(x;n-1)$. In Appendix C, we show that this form of
the joint distribution implies that 
%\red{What do you mean by the ``scaling form''?
  %Need a formal definition here}. 
the mean value of $\kappa^\ast = 2(1-\rho) =
2\langle\frac{\epsilon}{x_n+\epsilon}\rangle$, % using  
% the above distribution we show in the Appendix C that,
and in the  
large $n$ limit it is given by 
\[
 \kappa^\ast \sim n^{-1/\beta}.
\]
For % an exponential
a {uniform}
distribution  $\beta=1$, and $\kappa^\ast$ is inversely proportional to
$n$.
 The asymptotic decay  
of the transition point is fastest for this class of distributions.
Comparison of the result with numerical data in \figurename{~\ref{scaling}a} reveals an excellent match.

\vskip 5 pt
\noindent{\bf (ii) Unbounded distributions with light tails}~
Next, we consider the case where the support of $Q$ is unbounded; however, the
tail of $Q$ is faster than power law.  We restrict ourselves to distributions of
the form $Q(x)\sim\exp(-x^\delta)$,with $\delta>0$~\cite{satya}, which
cover a broad class of distributions that   
occur in natural settings, including exponential, Gaussian and stretched
exponential form~\cite{stretched}.  In this case, the limiting
distribution of the scaled maximum $z = b_n^{-1}(x_n-a_n)$ is the Gumbel
distribution of the form  
$f(z)=e^{-z-e^{-z}}$, and 
\begin{eqnarray}
  a_n&=& (\ln
         n)^{\frac{1}{\delta}}+\frac{1}{\delta}(\frac{1}{\delta}-1)(\ln
         n)^{\frac{1}{\delta}-1} \ln(\ln n) \\ 
  b_n &=& \frac{1}{\delta}(\ln n)^{\frac{1}{\delta}-1}.
\label{ab}
\end{eqnarray}

%{Clearly write an expression for $f(z)$ and the scaling constants
 % $a_n$ and $b_n$ for all the other cases if there are any. Done.}
% The maximum value is localized around $a_{n}$, and $b_n \to \infty$ if
% the distribution is slower  
% than an exponential and $b_n \to 0$ if it is faster than an
% exponential. In this case, 
% \[
% \bar{z} \approx {n}{}\int dx \int d\epsilon~\frac{\epsilon}{x}
%                    Q(x+\epsilon) P_{max}(x;n-1),
% \]
% and the joint distribution 
% \[
%  P(x,\epsilon) = n Q(x+\epsilon) P_{max} (x,N-1),
% \]
% where $\epsilon \ge 0$.
Using this limiting form, and a calculation similar to the bounded case (see
Appendix C for details), we get 
\[
 \kappa^\ast\sim {(\ln n)}^{-1}.
 \]
In this case, the coding transition point goes to zero very slowly. Consequently,
even in a highly nutrient diverse environment, % we cannot be sure that
the organism % system is in the 
may not sense and encode the environment. 
% coding regime. 
Comparison with numerical simulation data is shown in \figurename{~\ref{scaling}b}.
\vskip 5 pt

\noindent{\bf (iii) Power law distributions}~~
When $Q$ has a power law tail, we can show that $\kappa^\ast$ does not go
to zero in the large $n$ limit. 
Consider $Q(x)$ of the form $Q(x) = \alpha e^{-x^{-\alpha}} x^{-(1+\alpha)}$,
$\alpha > 0$. In this case, the asymptotic distribution    
$f(z) \sim \frac{\alpha}{z^{1+\alpha}} e^{-z^{-\alpha}}$ is the Fr{\'e}chet distribution, with $a_n=0$,
and 
$b_n=n^\frac{1}{\alpha}$. 
% \iffalse
% So, we have 
% \begin{eqnarray}
%  \bar{z} &=& \frac{n}{b_n} b_n^{1+\alpha}\int dx \int d\epsilon ~ 
%  \frac{\epsilon}{x+\epsilon}
%  \frac{e^{-(x+\epsilon)^{-\alpha}}}{(x+\epsilon)^{1+\alpha}}
%  x^{-(1+\alpha)}  
%  e^{-({\frac{x}{b_n}})^{-\alpha}} \nn \\
%  &=& {n} b_n^{\alpha} \int_0^\infty dx~x^{-(1+\alpha)}
%  e^{-({\frac{x}{b_n}})^{-\alpha}}  \int_0^\infty d\epsilon ~  
%  \frac{\epsilon e^{-(x+\epsilon)^{-\alpha}}}{(x+\epsilon)^{2+\alpha}}  \nn \\
% &\approx& {n} b_n^{\alpha} \int_0^\infty dx~x^{-(1+\alpha)}
% e^{-({\frac{x}{b_n}})^{-\alpha}}  \int_0^\infty d\epsilon ~  
%  \frac{\epsilon}{(x+\epsilon)^{2+\alpha}}  \nn \\
%  &=& {n} b_n^{\alpha} \int_0^\infty dx~x^{-(1+\alpha)}
%  e^{-({\frac{x}{b_n}})^{-\alpha}} x^{-\alpha}  
%  \int_0^\infty dy ~ 
%  \frac{y}{(1+y)^{2+\alpha}}  \nn \\
%  &\sim&  {n} b_n^{\alpha} \int_0^\infty dx~x^{-(1+\alpha)}
%  e^{-({\frac{x}{b_n}})^{-\alpha}} x^{-\alpha}\nn \\ 
%  &=& {n} b_n^{\alpha} \int_0^\infty dx~x^{-(1+\alpha)}
%  e^{-({\frac{x}{b_n}})^{-\alpha}} x^{-\alpha}\nn \\ 
%  &=& {n} b_n^{-\alpha}\int_0^\infty dw~w^{-(1+\alpha)}
%  e^{-({{w}})^{-\alpha}} w^{-\alpha}\\ 
%  &=& \int_0^\infty dw~w^{-(1+\alpha)}  e^{-({{w}})^{-\alpha}} w^{-\alpha}\\
%  &\sim& O(1),
% \end{eqnarray}
% which is a contradiction. 
% \fi 
%We % can calculate the ratio of highest to second highest values, and
%show that in this case the ratio of the largest to the second
%largest value does   
% that it does
%not go to $1$ as  $n\rightarrow \infty$. 
As before, let $x$ denote the
second largest value and $x+\epsilon$ denote the highest value. Then 
% If the highest value is
% $x+\epsilon$, and the second highest is $x$, then we can show  
% that in the limit of large $n$, 
% \begin{eqnarray}
%  \langle \frac{\epsilon}{x} \rangle &\sim& \frac{{n}^{\frac{2}{a}}}{a-1}
%  ~\mbox{for $a > 1$} \nn \\ 
%  &=& \infty ~\mbox{for $a \leq 1$.}
% \end{eqnarray}
% We show it as follows: 
\begin{eqnarray}
 {\langle \frac{\epsilon}{x} \rangle}  &\sim& \frac{n^{\frac{2}{1+\alpha}}}{\alpha-1}, \mbox{~ $\alpha > 1$}\nn\\
 &\sim& \infty, \mbox{~~~~~~~$\alpha \le 1$.}
\end{eqnarray}
The details of the calculation are in Appendix C. 
For $n \rightarrow \infty$,  we see that 
% We see that in the large n limit, 
the mean ratio of the highest to second highest values diverges, even for
$\alpha >1$ (where the mean of the  
parent distribution is finite). We show in Appendix C that this implies $\kappa^\ast$ also diverges
asymptotically. The heavy tails of  
the distributions cause the largest selected value to dominate over
all other values, and thus the organism is not in the coding regime.  
\subsection{Discussion} 
When % An organism in the regime
$\kappa>\kappa^\ast$, the organism senses the environment, % is engaged
% in a decision making process, and its
% response  
encodes information about the environment, and takes an appropriate
response. Note that the response $i$ is a potentially a random map from
the environment state $\alpha$. This stochastic response is, often,  critical for the
survival of the population. The value of $\kappa^\ast$ depends on the
distribution of environmental states and the free benefit associated the
particular response; therefore, the precise value of $\kappa^\ast$  cannot
be precisely calculated. However, universal 
behavior emerges in the  limit of a large number of environmental
states. The behavior falls into three different classes % s, which are  
% consistent with the hierarchy of the behavior
as a function of the tail of the 
distributions of the environment frequency $f_{\alpha}$ and the free
energy benefit $E_{\alpha}$. We show that only limiting behaviour of
$\kappa^\ast$ is a function of the product $g_{\alpha} = f_{\alpha}
E_{\alpha}$. 
When $g_{\alpha}$ are chosen from a bounded distribution, the mean $\kappa^\ast$
goes to $0$  as $n^{-1/\beta}$ as a function of the number of environment
states $n$.   
%In the ecology literature, s
%such organisms are often dubbed {\it generalists} \cite{generalists}, which can use a number of different nutrient sources for its metabolic needs. 
In the case of unbounded distributions with light tails, the mean $\kappa^\ast$
decays to zero  as $(\ln n)^{-1}$, and the mean $\kappa^\ast$ can diverge
for 
% the parent distribution has finite moments,
% whereas it diverges for 
power law % parent
distributions. The divergence is an artifact of the power law
distribution, but what it does imply is that there  
is no typical $\kappa$ above which we can expect an organism to be in the
coding regime.  
The slow convergence in the 
case of  exponential or Gaussian tails is more surprising,
since one would have expected 
the organism to code for the environment % regime for a large number of
                                % nutrients
when the product is drawn from such distributions. 

Which of these cases, if any, occur in natural environments is a difficult
question to answer. We may expect that in highly  
biodiverse environments, such as in soil microbial communities, the number
of available nutrients is high.  
Such communities harbor hundreds of different species of microbes, and
their interactions  
often promote chemical diversity \cite{metabolites}. 
A thorough quantitative study of nutrient distributions in such
communities would  
be helpful in making predictions about the coding behavior of metabolic systems.

\section{Conclusion}
We have formulated the problem of adaptive response in biological systems
as an optimization problem where the  tradeoff is   
between the benefit obtained from specific responses and the cost of
producing them. Our main result   
is that there is a cascade of responses as a function of the cost of
information, and the number of  
responses that are active in an optimal metabolic system depends on the
cost and the distribution of nutrients.  
Even when the system is capable of producing $n$ different responses, the
number of responses it actually uses can be significantly smaller.  

The decision to express certain genes is executed by the regulatory
machinery, and evolutionary processes may lead  
to machinery that produces the optimal response when supplied with a
certain environment.  
The traditional  
view has been that in order to metabolize new nutrients, genotypic changes
that produce new or altered enzymes are required.  
However, our model points to the  
existence of 
epigenetic mechanisms that carry out a decision theoretic task -- that of
constructing the optimal response channel 
from within the repertoire of responses already available. 
%Recent years have seen an explosion in the field of epigenetics, and indeed regulatory mechanisms that cause switches 
%between metabolic phenotypes are now known in some detail 
%For example, a prion protein like element converts wild fungi from metabolic specialists
%to generalists \cite{fungi}, pathogenic fungi can switch from biotrophic to necrotrophic lifetyles \cite{biotrophic}, etc. 
Our model is based on a particular way of looking at such regulatory
mechanisms -- namely that they % evolved to solve a class
optimal solutions of a particular % problems
in rate distortion problem.

In this % present
paper, we only characterize  the structure of the optimal channe. We %  have
% made no hypotheses
do not describe how these regulatory  
mechanisms evolve, or indeed about the kind of molecular processes
necessary to establish the optimal channel.  
More detailed studies are required to address these questions. Further
investigations may bring out new surprises, 
but we believe that some of the central features of our results will
continue to hold.

% \iffalse
% We have seen that as the number of nutrients in the system grows, the
% coding transition point shows three different kinds of  
% behavior. Only in one case -- where the expected benefit values are
% drawn from a bounded distribution -- the transition  
% point goes to zero quickly as the number of nutrients grows. It is
% reasonable to ask what this implies about molecular coding  
% in the metabolic world.  

% Strictly speaking, the results here apply if the number of nutrients
% available to a cell is sufficiently large.  
% However, in the case where the free energy benefits are drawn from a
% bounded distribution, we see that the asymptotic behavior  
% already holds for $n \sim 10$ (see \gurename{ \ref{scaling}}a). When the
% benefits are drawn from an unbounded  
% distribution, the approach to zero either does not happen or is
% logarithmically slow, and we cannot conclude that  
% the cell is generically in the coding regime. 

% It is likely that environments corresponding to all the different
% regimes exist. Extreme metabolic specialists may inhabit  
% environments where the beneifts follow the power law distribution, such
% that a single nutrient dominates the scenarion. 
% \fi

\section{Appendix}

\subsection{Algorithm for finding optimal channel}
Here we consider the problem of computing the optimal $p(i\vert \alpha)$
for $\kappa \neq 0$. Let
\begin{eqnarray*}
J(p,q) & = & \sum_{i\alpha} f_{\alpha} p(i\vert \alpha) \log \left(
  \frac{p(i\vert \alpha)}{q(i)} \right)\\
  & = & I(p) + D(p\|q),
\end{eqnarray*}
where $D(p,q)$ denotes the relative entropy of $p$ and $q$. 
Then, $J(p,q)$ is jointly convex in $(p,q)$ and moreover, $\min_{q} J(p,q)
= I(p) + \min_q D(p,q) = I(p)$. 
Since $G(p) - \kappa^{-1} J(p,q)$ is jointly concave in $(p,q)$, and
$\max_{(p,q)} G(p) - \kappa^{-1} J(p,q) = \max_p G(p) - \kappa^{-1} I(p)$,
it follows
that the coordinate descent will converge to the optimal solution. Thus,
the following algorithm will compute the optimal solution for
$\kappa>0$:
\begin{enumerate}[1.]
\item Set $q^0 = \frac{1}{m}$ where $m$ is number of responses.
\item % Compute the solution of the optimization problem
  For $k \geq 0$, set
  \[
  p^{k} \leftarrow \argmax_{p} \{ G(p) - \kappa^{-1} J(p,q^k)\}
  \]
  It is easy to check that $p^k(i\vert \alpha) \propto q^k(i) e^{\kappa
    E_{i\alpha}}$. Thus, $p^k(i\vert \alpha) = \frac{q^k(i) e^{\kappa
    E_{i\alpha}}}{\sum_j q^k(j) e^{\kappa E_{j\alpha}}}$
\item % Compute the solution of the optimization problem
  For $k \geq 0$, set
  \[
    q^{k+1} \leftarrow \max_{q} \{ G(p^k) - \kappa^{-1} J(p^k,q)\}.
  \]
  It is easy to check $q^{k+1}(i) = \sum_{\alpha}  f_{\alpha} p^k(i\vert \alpha)$.
% \item Take $q(i)=q^\ast(i)$, $p(i \vert \alpha)= p^\ast(i\vert \alpha) $
% and go to step 1. 
\end{enumerate}
This algorithm converges to the optimum solution. 
\subsection{Product of two random variables}
Suppose $f_i$, $i = 1, \ldots, n$ are IID samples from $P_f$ and $E_i$, $i
= 1, \ldots, n$, are IID samples from $P_e$. We assume that $P_e$ is
supported over $[0,a]$ with $a < \infty$. 
 Let $g_i = f_i E_i$, $i = 1,
\ldots, n$, and let its distribution be denoted by $P_g$. Permute indices such that $g_1 = f_1E_1$ is the largest value
among $\{g_i\}$ and $g_2 = f_2E_2$ is the second larges value among
$\{g_i\}$. We will restrict ourselves to distributions such that $g_2/g_1
\rightarrow 1$ as $n \rightarrow \infty$.
Recall that $s_i = \frac{E_i}{E_1}$ and $s_iw_i = \frac{g_i}{g_1} =
\frac{f_iE_i}{f_1E_1} \leq 1$. Note that $s_2w_2 \rightarrow 1$ as $n
\rightarrow \infty$.

% There are $n$ such equations.
Let $x_i>1$ denote the non-trivial solution of \eqref{eq:wi} corresponding
to $w_i$, $i = 2, \ldots, n$. Let $y_i = x_i -1$. 
% Let the smallest value in the set
% $\{y_i\}$ be 
% $y_t$. We restrict ourselves to parent distributions 
% such that $y_t \ll 1$ for large $n$. \red{Is it clear, apriori, that such
%   distributions exist?} Our goal is to find $y_t$ for large
% $n$.   % Consider any $y_i \ll 1$ (this includes the case 
% $y_t$). \red{Why does such a $y_i$ exist?}
Substituting $1+y_i$ in \eqref{eq:wi}, and expanding up to second order in
$y_i$, we obtain the solution 
\begin{equation}
  y_i= \frac{2(1-s_i w_i)}{1-\frac{1}{2} s_i w_i (1-s_i)}.
  \label{eq:smally}
\end{equation}
This solution is valid provided $y_i$ is small in the limit of large
$n$. Since $s_i \geq 0$ and $s_iw_i \leq 1$, it follows that 
\[
y_i \leq 4(1-s_iw_i).
\]
Since $s_2w_2 \rightarrow 1$, it follows that $y_2 \rightarrow 0$. Since
the coding transition is determined by the smallest value of $x_i$, $i
\geq 2$, in the limit of large $n$, one can limit to indices $i$ with $y_i
\ll 1$.

% Since $y_i$ is small (which was used to do the expansion above), the above
% expression must be small.  
Let $A_i = 2(1-s_iw_i)$ denote the numerator  and $B_i = 1 -
\frac{1}{2}s_iw_i(1-s_i)$ denote the denominator in the expression of
$y_i$. Then we have that $0 \leq A_i \leq 2$ and $\frac{1}{2} \leq B_i \leq 1 +
\frac{1}{2}s_i$. We show below that there exists a bound $S$ such that, in
the limit of large $n$, $s_i \leq S$ for all $i \geq 2$. Thus, $B_i \leq
1 + S/2$ in the limit of large
$n$. Thus, $y_i \ll 1$ only if $s_iw_i \lesssim 1$. We also show that $s_i
\lesssim 1$ for all $i$ such that $s_iw_i \lesssim 1$. Thus, it follows
that $y_i \approx 2(1-s_iw_i)$.

% Notice that the numerator N is bounded above by 
% $N=2$ (since $s_i w_i \le 1$), and the denominator $D$ is bounded below
% by $1/2$ (since $s_i w_i \le 1$ and $s_i > 0$). Thus,  
% $y_i$ is small in one of three ways: either $N$ is small but $D$ is not
% large, or $D$ is large but $N$ is not small,  
% or both $N$ is small and $D$  
% is large. (The case where both $N$  and $D$ vanish or both diverge but
% their ratio is small is ruled out, since $N$ is bounded  
% above and $D$ is bounded below).

% We notice that 
% \[
% D=1+\frac{1}{2}(s_i w_i) s_i-\frac{1}{2}s_i w_i \leq 1+\frac{1}{2}(s_i
% w_i) s_i \leq    
% 1+\frac{1}{2}s_i.\]
% At this point, we introduce a condition: $E$ is chosen from a bounded distribution.
% We show below that when this is the case, $s_i \not \gg 1$, and
% therefore $D$ is not large. We are then left with the only  
% option that $N$ must be small, which means that $s_i w_i \lesssim 1$. This condition holds for all $y_i \ll 1$. 
%From this, we 
%see that $y_t$ corresponds to the highest value of $s_i w_i$. 

% We show one further thing below, which is that when $E$ is bounded, then
% for all $i$ such that  
% $s_i w_i \lesssim 1$, we must have $s_i \lesssim 1$. Then we have 
% $y_i = (1-s_i w_i)$ (which holds for all $y_i \ll 1$). From this, it becomes clear that 
% $y_t$, which is the smallest $y_i$, corresponds to the largest value of $s_i w_i$.

% We need to complete the proof by showing that when $E_i$ is bounded, the
% following two statements hold:\\  
\begin{lemma} Suppose the distribution $P_e$ is supported over the
  bounded set $[0,a]$. Then the following holds:
  \begin{enumerate}
  \item there exists a bound $S$ such that $s_i =
    \frac{f_i}{f_1} \leq S$ for all $i \geq 2$ and sufficiently large $n$.
  \item Suppose $s_i w_i \lesssim 1$. Then we must have $s_i \lesssim 1$.\\
  \end{enumerate}
\end{lemma}
\noindent We break our proof down
into two cases.

\texitem{(1)} $P_f$ is bounded with support on $[0,c]$. \vskip 5pt

% \noindent{\it  Proof of statement (1)} \vskip 5 pt

\noindent In this case,
$P_g$ is supported on $[0,ac]$. Since $P_g$ has finite support,  
it belongs to the Weibull domain, by definition. In this case, the
distribution of $g_1$ is $b_n^{-1} h(\frac{g_1-ac}{b_n})$, where 
\[
h(z)\sim{(-z)}^{\beta-1} \exp\big(-(-z)^\beta\big)\ 
\] 
and $b_n = n^{-1/\beta} \ll a$.  This distribution is concentrated at 
$g_1 \lesssim ac$. 
%Furthermore, for values drawn from a
%Weibull domain distribution, $g_2 \lesssim g_1$.  
% Consider the particular
% form $P_e(E) \sim (a-E)^{\beta-1}$,    
% $\beta>0$ with the support $0 \le f\le a$. (\red {This form is probably
%   not essential. But we can discuss that if you find the  
% proof of this special case acceptable.})
% and $Q(x)=0$  
% for $x > a$. Also, $\beta>0$. 
Since $f_1 \leq c$ and $E_1 \leq a$, % we see that their
the product $g_1 = f_1 E_1
\lesssim ac$ if, and only if,   
$f_1 \lesssim c$ and $E_1 \lesssim a$. The latter condition means that
$E_i/E_1 \leq a/E_1 \rightarrow 1$ as $n \to \infty$.  Thus, we are
guaranteed that the first statement in the Lemma holds for any $S>1$.
%does not hold for any $i$, i.e $s_i \not \gg 1$. \\ \vskip 2 pt
% \noindent {\it Proof of statement (2)} \\ \vskip 2 pt

Since the denominator in \eqref{eq:smally} is bounded, $y_i$ is 
small only if $s_i w_i \lesssim 1$, or equivalently, $g_i \lesssim g_1$.
Since $g_1 \lesssim ac$, it follows that 
% Furthermore, for values drawn from a
% Weibull domain distribution, $g_2 \lesssim g_1$,  
% implying that 
$g_i \lesssim ac$. This holds only if $f_i \lesssim c$ and $E_i
\lesssim a$, i.e. $s_i = E_i/E_1 \approx 1$. This establishes the second
statement in the Lemma.
\vskip 5 pt

\texitem{(2)} $P_f$ % is an 
is unbounded % distribution
with a light tail. \vskip 5pt

\noindent We consider $P_f(f) \sim \exp(-f^\delta)$ 
with $\delta>0$. Thus, 
\begin{eqnarray*}
\lefteqn{P(g)}\\
& = & \int_{a^{-1}g}^\infty  f^{-1} P_f(f) P_e(f^{-1} g)\ dx \\
 & = &  \int_0^\infty 
    (a^{-1}g+\epsilon)^{-1}P_f(a^{-1}g+\epsilon)
      P_e\big(\frac{g}{a^{-1}g+\epsilon}\big)~d\epsilon       
\end{eqnarray*}
We are interested in the tail of the distribution, i.e. $P(g)$ for $g \gg
1$. 
%\red{What about other unbounded distributions? What happens in those
%  cases?}
For large  
$g$, % we can make the following approximation in the integrand: 
$P_f(a^{-1}z+\epsilon) \sim \exp(-(a^{-1}g)^\delta)\exp(-\frac{\delta
  \epsilon}{a^{-1}g})$. Then,
\begin{eqnarray*}
  &&P(g) \nn \\
  &\approx& e^{-(a^{-1}g)^\delta} \int_0^\infty \frac{\exp(-\frac{\delta
             \epsilon}{a^{-1}g})}{1+\frac{\epsilon}{a^{-1}g}}P_e\big(\frac{a}{1+\frac{\epsilon}{a^{-1}g}}\big)
             ~\frac{d\epsilon}{a^{-1}g}\\
      &\approx& P_f(a^{-1}g) \int_0^\infty \frac{\exp{(-\delta
             \epsilon^\prime)}}{1+\epsilon^\prime}P_e\big(\frac{a}{1+\epsilon^\prime}\big) ~d\epsilon^\prime \nn
  \\ 
      &\sim& P_f(a^{-1}g).\nn \\
\end{eqnarray*}
Thus, the tail of $P_g$ follows the same form as $P_f$, except
that $g$ is now scaled by $a^{-1}$. 

The distribution $P_f(f)$ belongs to
the Gumbel domain.  The distribution of the extreme value $f_{max}$ is $d_n^{-1}
h(\frac{f_{max}-c_n}{d_n})$, where $h(z)\sim \exp(-z-e^{-z})$,  
$c_n \sim (\ln(n))^{1/\delta}$, and $d_n \sim (\ln(n))^{1/\delta-1} \ll
c_n$. Therefore,  $f_{max}$ is localized around $c_n$, i.e $f_{max}/c_n \approx
1$.   
% For the parent distribution
The distribution $P(g) \sim P_f(a^{-1}f)$, so we have $g_1/(a c_n) \approx 1$.
Thus, $E_1 \approx (a c_n)/f_1$; however, since % the distribution of the
% maximum value of $f$ is localized around 
$f_1 \leq f_{max}$, and the maximum value of $E_1$ is $a$, this equality can be true
only if $c_n/f_1 \approx 1$ and $E_1 \approx a$. The latter condition 
implies that $E_i/E_1 \leq a/E_1 \to 1$ as $n \to \infty$. Thus, any $S>1$
satisfies the first statement in the Lemma.

% \\ \vskip 5 pt
% %does not hold for any $i$, ie $s_i \not \gg 1$. \\ \vskip 2 pt
% %For the Gumbel domain, we have 
% %$g_2/g_1 \to 1$. 

% \noindent {\it Proof of statement (2)} \\ 

% \noindent

Since the denominator in \eqref{eq:smally} is bounded, $y_i$ is 
small only if $s_i w_i \lesssim 1$, or equivalently, $g_i \lesssim g_1$.
% For small $y_i$ we must have $s_i w_i \lesssim 1$, which implies $g_i
% \lesssim g_1$, and therefore
Thus,  
$g_i/(a c _n) \approx 1$;  consequently, $E_i
\approx (a c_n)/f_1$, and because $f_1/c_n \approx 1$, we have  
$E_i \approx a$. % Comparing with $E_a$, we see that
Thus, it follows that $s_i = E_i/E_1 \approx 1$.

\subsection{$\kappa^\ast$ in different asymptotic regimes}
Suppose $X_i$, $i = 1, \ldots, n$, denote $n$ IID draws from the
density $Q$. Let $x$ denote the \emph{second} largest of the $n$ terms, and
let $x+\epsilon$ denote the largest term of the $n$ terms. Then the joint
density $P(x,\epsilon)$ of $(x,\epsilon)$ is given by $n
P(x,\epsilon) = Q(x+\epsilon)P_{\max}(x;n-1)$, where $P_{\max}(x;n-1)$ denotes the density
of the largest of $n-1$ draws from the distribution $Q$. We analyze the
behavior of the mean $\langle \kappa^\ast \rangle$ for three different
asymptotic regimes.

\noindent {\bf Bounded distribution:}\\
\begin{eqnarray}
\lefteqn{ \langle \kappa^\ast\rangle = 2 \Big\langle
  \frac{\epsilon}{x+\epsilon} \Big\rangle}  \nn \\
&\sim& {n}{}\int dx \int d\epsilon~\frac{\epsilon}{x+\epsilon} Q(x+\epsilon) P_{max}(x;n-1)\nn \\
&\approx& \frac{n}{a}\int dx \int d\epsilon~{\epsilon} Q(x+\epsilon) P_{max}(x;n-1)\label{case2:1} \\
&=& \frac{n}{a b_n}\int_0^a dx \int_0^{a-x} d\epsilon~{\epsilon}
    (a-x-\epsilon)^{\beta-1} e^{-{(\frac{a-x}{b_n})}^\beta} \nn \\
&=& \frac{n}{a b_n} b_n^\beta \int_0^{a/b_n} dz \int_0^{b_n z} d\epsilon~ \epsilon {\big(z-\frac{\epsilon}{b_n}\big)}^{\beta-1} 
z^{{\beta}-1} e^{-z^\beta}  \nn \\
&& \mbox{(with the transformation $z=\frac{a-x}{b_n}$)} \nn \\
%&& \text{\red{ Did you also
%   take a limit in $n$?}} \nn\\
&=& \frac{n}{a}~ {b_n}^{\beta+1} \int_0^{a/b_n} dz \int_0^z dy ~ y (z-y)^{\beta-1} z^{\beta-1} e^{-z^\beta} \nn \\
&& \mbox{(with the transformation $y=\frac{\epsilon}{b_n}$)} \nn \\
&\approx& \frac{n}{a}~ {b_n}^{\beta+1} \int_0^\infty dz \int_0^z dy ~y(z-y)^{\beta-1} z^{\beta-1} e^{-z^\beta} \label{case2:2} \\
&\sim& n^{-\frac{1}{\beta}}, \label{case2:3}
\end{eqnarray}
where \eqref{case2:1} follows from the fact that the integrand is non-zero only if $x\lesssim z$, \eqref{case2:2} 
follows from the fact that $b_n \to 0$ and the integrand vanishes at large
$z$, and \eqref{case2:3} follows  
from $b_n \sim n^{-\frac{1}{\beta}}$. 
\\

\noindent {\bf Unbounded distribution with light tails:}\\
\noindent With the scaling transformation $z=\frac{x-a_n}{b_n}$, we have 
\[
 P(z,\epsilon)=n Q(a_n + b_n z +\epsilon) f(z)
\]
\begin{eqnarray*}
  \lefteqn{\langle \kappa^\ast \rangle} \nn\\
& = & \int dz~\int ~ d\epsilon \frac{2\epsilon}{a_n+b_n
   z+\epsilon} e^{-{(a_n 
    + b_n z + \epsilon)}^\delta-z-e^{-z}}\nn \\ 
\end{eqnarray*}
Since $a_n \gg 1$, and $a_n \gg b_n$, the integrand has
finite values only when $z \sim O(1)$, and  
$\epsilon \ll a_n + b_nz$. Expanding the first term in the exponent as 
${(a_n+\epsilon)}^\delta +\delta(a_n+\epsilon)^{\delta-1}b_n z$, and using
the explicit forms of the scaling constants above, and the fact that
$\epsilon \ll a_n$, it follows that the second term is  $~c z$, where $c$
is independent of $n$. Thus, the integral over $z$ yields  
\[  
\langle \kappa^\ast \rangle \sim \frac{1}{n} \int d\epsilon~ \frac{\epsilon
  \exp[-(a_n+\epsilon)^\delta]}{a_n} 
\]
Using $\epsilon \ll a_n$, 
\begin{eqnarray}
  \langle \kappa^\ast \rangle &\sim& \frac{1}{n} {\exp(-a_n^\delta)}\int d\epsilon ~\epsilon
                \frac{\exp(-\delta a_n^{\delta-1}\epsilon)}{a_n} \nn \\ 
         &\sim& \frac{1}{n} \frac{\exp{(-\ln n -\ln ((\ln
                n)^{\frac{1}{\delta}-1}}))}{ a_n^{2\delta-1}}\nn \\ 
         &\sim& \frac{1}{b_n a_n^{2\delta-1}}\nn \\
         &\sim& \frac{b_n}{a_n}\nn \\
         &\sim& \frac{1}{\delta} {(\ln n)}^{-1}
\end{eqnarray}
where the last two steps use the forms of the scaling constants given
in \eqref{ab}.\\
%Therefore,  
%\[
% \kappa^\ast\sim {(\ln n)}^{-1}.
% \]

\noindent {\bf Power Law:}
% \red{$\langle \kappa^\ast \rangle \sim \Big\langle \frac{\epsilon}{x+\epsilon}
% \Big\rangle$. In the power law case $\epsilon \gg 1$, and so the behavior
% of $\Big\langle \frac{\epsilon}{x+\epsilon}
% \Big\rangle$ could potentially be very different from $\Big\langle \frac{\epsilon}{x}\Big\langle$. 
% In fact, since $\epsilon \gg 1$, it might be the case that $\Big\langle \frac{\epsilon}{x+\epsilon}
% \Big\rangle \sim 1$. Can you comment?}\\
%\iffalse
% \iffalse
% Since we have $s_i w_i \ll 1$, it must be true that $s_i \ll 1$ or $w_i \ll 1$.
% First, consider the case $w_i \ll 1$. 
% Rearrange \eqref{eq:wi} to get 
% \begin{equation}
%  w_i = \frac{x-1}{x(x^{s_i}-1)}%=\frac{x-1}{x(x-1)}=\frac{1}{x}.
% \label{eq:wi2}
%  \end{equation}
% Then \eqref{eq:wi2} can be satisfied only if $x \gg 1$, and therefore $\kappa$ is large. 
% Consider the alternate possibility, $s_i \ll 1$. Consider the case that
% $x$ is {\it not} large. Then, $x^{s_i} \approx 1$,  
% and then $w_i \approx 1+w_i -\frac{1}{x}$, which is solved by $x \approx
% 1$, which we ruled out above. Thus, $x$ must be large. 
% \fi 
% \iffalse
% Consider the other case, $s_i \ll 1$. Once again, the only way in which
% $w_i \ll 1$ is if $x \gg 1$ and so $\kappa \gg 1$.}  
% If $E_i$ is chosen from a bounded distribution, 
% then the only way $g_1 \gg g_2$ is if $f_1 \gg f_2$. Then $s_i$ is bounded above whereas 
% $w_i \approx 0$; using this in \eqref{eq:wi}, we see that for finite
% $x$, the lhs goes to $0$ and the rhs goes to $-1$. The  
% only way to avoid this is if $x$ diverges, and therefore $\kappa^\ast$ diverges.
% \fi
The ratio of largest to second-largest value is $1+\langle \frac{\epsilon}{x}\rangle$. Now,
\begin{eqnarray*}
  \lefteqn{\langle \frac{\epsilon}{x} \rangle} \nn \\
  &=& n^{} \int dx~ P_{max}(x,n-1) \int d\epsilon \frac{\epsilon}{x}
      Q(x+\epsilon)\nn  \\ 
  &\sim& n \int dx~ P_{max}(x,n-1) \frac{x^{1-\alpha}}{\alpha-1} \label{pl1}\\
  &=& \frac{n}{\alpha-1} b_{n-1}^\alpha\int dx
     ~ {x^{-2\alpha}}e^{-\big(\frac{b_{n-1}}{x}\big)^\alpha}\\ 
  &\approx& \frac{n^{\frac{2}{\alpha}}}{\alpha} \int dy ~ y^{-2\alpha}
            e^{-\big(\frac{1}{y}\big)^{\alpha}} \nn \\ 
  &\sim&\frac{n^{\frac{2}{1+\alpha}}}{\alpha-1}.\nn
 \end{eqnarray*}
% \noindent {\bf Response }
Therefore the ratio of largest to second largest value diverges, i.e $g_1 \gg g_2$. 

Since $\kappa^\ast$ is not small, the formula 
$\langle \kappa^\ast \rangle \sim \Big\langle \frac{\epsilon}{x+\epsilon}
\Big\rangle$ does not work. We will demonstrate now that $\kappa^\ast$, in
fact, diverges.  
%Assuming it does, the ratio is $\frac{x}{x+\epsilon}$, not
%$\frac{\epsilon}{x+\epsilon}$.  
%The former 
%goes to $0$, not $1$. 

The equation \eqref{eq:wi} is exact. 
Rearrange \eqref{eq:wi} to get 
\begin{equation}
 s_i w_i= \frac{s_i(x-1)}{x(x^{s_i}-1)}.%=\frac{x-1}{x(x-1)}=\frac{1}{x}.
\label{eq:wi2}
 \end{equation}
We have seen that for power law, the largest term $g_1 \gg g_2$, the
second largest term, and therefore $s_i w_i =
\frac{g_i}{g_1} \rightarrow 0$ for all $i$. We show that this is  
possible only if $x \gg 1$, and therefore $\kappa \gg 1$. 

First, it is easy to show that for $x > 1$ (which is our region of interest) 
\[
s_i w_i =\frac{s_i(x-1)}{x(x^{s_i}-1)} \ge \frac{1}{x^{1+s_i}}.
\]
Since $s_i w_i \rightarrow 0$, we must have $\frac{1}{x^{1+s_i}}
\rightarrow 0$.
%Consider the case where $E_i=E$ for all $i$, and therefore $s_i=1$. Then
%we require $w_i \ll 1$. 
% \iffalse 
% First, consider the possibility that $x=1+a$, where $a \ll 1$, which
% corresponds to $\kappa \approx 0$. Putting this in  
% \eqref{eq:wi2}, and using the expansion $(x^{s_i}-1)={(1+a)}^{s_i}
% -1\approx s_i a$, we have  
% \begin{equation} 
% s_i w_i\approx \frac{1}{x}.  
% \end{equation}
% Since $g_1 \gg g_2$, we must have $s_i w_i \ll 1$. Therefore, the above
% solution cannot hold, since $x \approx 1$  
% by assumption. Thus we have stablished that $x$ cannot be close to $1$,
% and equivalently $\kappa$ cannot be small.  
% Since $ s_i w_i \ll 1$, the rhs of \eqref{eq:wi2} must be small. We show
% in the following that this would require  
% $x \gg 1$, and therefore $\kappa \gg 1$.
% Suppose $s_i$ is finite. In that case, since $x$ is not close to $1$, 
% the rhs small only if $x\gg 1$. Suppose $s_i$ is small; then it is
% easily shown that the rhs goes to the value  
% $(x-1)/(x \ln x)$. Once again, this can be small only if $x \gg 1$,
% since $x$ is not close to $1$. 
We show below that $s_i$ is bounded with high probability, and therefore,
we must have $x \rightarrow \infty$ with high probability. 

% in order to have $s_i w_i \ll 1$.  
%Finally, the rhs can be small if $s_i \gg 1$, but we rule out this possibility below.

As before, consider the case that $E$ is drawn from a bounded
distribution. Further, let $f$ be drawn from 
a power law distribution, $P(f) \sim \frac{\exp(-f^{-\alpha})}{f^{1+\alpha}}$. Then
\begin{eqnarray*}
\lefteqn{P(g)}\\
& = & \int_{a^{-1}g}^\infty  f^{-1} P_f(f) P_e(f^{-1} g)\ dx \\
 & = &  \int_0^\infty 
    (a^{-1}g+\epsilon)^{-1}P_f(a^{-1}g+\epsilon)
      P_e\big(\frac{g}{a^{-1}g+\epsilon}\big)~d\epsilon\nn \\     
  &\approx&\int_0^\infty \frac{1}{(a^{-1}g+\epsilon)^{2+\alpha}} P_e\big(\frac{a}{1+\frac{\epsilon}{a^{-1}g}}\big)
             ~\frac{d\epsilon}{a^{-1}g}\\
               &\approx& \frac{1}{(a^{-1}g)^{2+\alpha}}\int_0^\infty \frac{1}{(1+\frac{\epsilon}{a^{-1}g})^{2+\alpha}} P_e\big(\frac{a}{1+\frac{\epsilon}{a^{-1}g}}\big)
             ~{d\epsilon}\\
               &\approx& \frac{1}{(a^{-1}g)^{1+\alpha}}\int_0^\infty \frac{1}{(1+\frac{\epsilon}{a^{-1}g})^{2+\alpha}} P_e\big(\frac{a}{1+\frac{\epsilon}{a^{-1}g}}\big)
             ~\frac{d\epsilon}{a^{-1}g}\\
      &\approx& P_f(a^{-1}g) \int_0^\infty \frac{1}{(1+\epsilon^\prime)^{2+\alpha}}
      P_e\big(\frac{a}{1+\epsilon^\prime}\big) ~d\epsilon^\prime \nn
  \\ 
      &\sim& P_f(a^{-1}g).\nn 
\end{eqnarray*}
Since the tail of $P_g$ behaves similar to that of $P_f$, it follows that
$P(g) \sim {g}^{-(1+\alpha)}$, for $g \gg 1$. Next, 
\begin{eqnarray}
 {P}(E|g) &=& \frac{P(g|E) P_e(E)}{P(g)}\nn \\
 &=& \frac{P_f(\frac{g}{E})P_e(E)}{P(g)}\frac{df}{dg} \label{eq:eg}\\
  &\sim& E^{\alpha} P_e(E) \nn \\ \nn
% &\sim& \frac{E^{1+\alpha}}{{g^{1+\alpha}}}\nn \\. 
\end{eqnarray}
where we have used that $g \gg E$ and used the forms of the tails of $P_f$ and $P_g$ in the last
step. Note that the $P(E\vert g)$ is independent of $g$ in the limit of 
% We see that the rhs is independent of $g$ (note that this happens only for
$g \gg 1$ and $g \gg E$.
%and so we simply denote $P_c(E)=P(E|g)\sim E^{1+\alpha} P_e(E)$, for large $g$.
%So $P(E_1|g_1)\sim \frac{E_1^{1+\alpha}}{{g_1^{1+\alpha}}}$. 
Thus, 
\begin{eqnarray*}
\lim_{\epsilon \rightarrow 0} P(E < \epsilon| g) & = & \lim_{\epsilon
                                                       \rightarrow 0}
                                                       \int_0^\epsilon ~dE ~ {E^{\alpha}}P_e(E)\\
  & \leq & \lim_{\epsilon \rightarrow 0} \Big(\epsilon^{\alpha} \int_0^\epsilon ~dE~ P_e(E)\Big) = 0.
\end{eqnarray*}
%  &\ll& \int_0^\epsilon ~dE~ P_e(E),
%  %\sim \epsilon^{2+\alpha} \to 0
% \end{eqnarray}
% where the last relation holds for $\epsilon \ll 1$.
Fix $\delta \ll 1$. Then, there exists $\epsilon>0$, independent of $n$, such that $P(E <
\epsilon| g) < \delta$ when $g \gg 1$. 
% Since the last term above is $\le 1$, we see that
% $P(E< \epsilon|g ) \ll 1$, and therefore with very very high probability,
% $E \not \ll 1$ for sufficiently large   
% $g$. 
Recall that in the power law case, the largest term $g_1 \gg 1$, it
follows that, with probability $1-\delta$, 
%  we must have $E_1 \not \ll 1$, or
% equivalently $\frac{1}{E_1} \not \gg 1$. Further, $E_i \le a$; therefore,  
$s_i = E_i/E_1 \le \frac{a}{\epsilon}$ for all $i$ and $n \geq 1$. 
% , implying that $s_i \not \gg 1$.   

% \iffalse
% \begin{eqnarray}
%  P(E_1=\epsilon_1 | g_1) &=& P(g_1 | E_1=\epsilon_1) \frac{P(E_1=\epsilon_1)}{P(g_1)}\nn \\
%  &=&P(g_1 | E_1=\epsilon_1) P(E_1=\epsilon_1) g_1^{1+\alpha}\nn \\
%   &=&P(g_1,\epsilon_1) g_1^{1+\alpha}\nn
% % &=& P(f_1=
% \end{eqnarray}
% \fi

%Following  arguments 
%similar to those in the previous subsection, we see that %$g_i \sim f_i a$ for large $g_i$, and therefore
%$s_i \not \gg 1$.

\end{document}